\documentclass[10pt, conference, compsocconf]{IEEEtran}\usepackage[utf8]{inputenc}
\usepackage{graphicx}
\usepackage{caption}
\usepackage{mathtools}
\usepackage{amsmath}
\usepackage{amssymb}
\usepackage{tabularx}
\usepackage[utf8]{inputenc}
\usepackage{lipsum,subcaption}
\usepackage[draft]{todonotes}% notes showed

\DeclarePairedDelimiter\floor{\lfloor}{\rfloor}
\begin{document}

\title{Avoiding Communication in Proximal Methods for Convex Optimization Problems}

\author{\IEEEauthorblockN{Saeed Soori\IEEEauthorrefmark{1},
Aditya Devarakonda\IEEEauthorrefmark{2},
James Demmel\IEEEauthorrefmark{2}, 
Mert Gurbuzbalaban\IEEEauthorrefmark{3} and
Maryam Mehri Dehnavi\IEEEauthorrefmark{1}}
\IEEEauthorblockA{\IEEEauthorrefmark{1}School of Electrical and Computer Engineering, 
Rutgers University, Piscataway, NJ\\}
\IEEEauthorblockA{\IEEEauthorrefmark{2}EECS Department, University of Berkeley, Berkeley, CA\\}
\IEEEauthorblockA{\IEEEauthorrefmark{3}MSIS Department, Rutgers University, Piscataway, NJ\\}
saeed.soori@rutgers.edu, (aditya, demmel)@eecs.berkeley.edu, (mert.gurbuzbalaban, maryam.mehri)@rutgers.edu}

\maketitle
%\section{Abstract}

\setcounter{page}{1}

\begin{abstract}
The fast iterative soft thresholding algorithm (FISTA) is used to solve convex regularized optimization problems in machine learning. Distributed implementations of the algorithm  have become popular since they enable the analysis of large datasets. However, existing formulations of FISTA communicate data at every iteration which reduces its performance on modern distributed architectures. The communication costs of FISTA, including bandwidth and latency costs, is closely tied to the mathematical formulation of the algorithm.
This work reformulates FISTA to communicate data at every “\textit{k}” iterations and reduce data communication when operating on large data sets. 
We formulate the algorithm for two different optimization methods on the Lasso problem and show that the latency cost is reduced by a factor of \textit{k} while bandwidth and floating-point operation costs remain the same. The convergence rates and stability properties of the reformulated algorithms are similar to the standard formulations. The performance of communication-avoiding FISTA and Proximal Newton methods is evaluated on 1 to 1024 nodes for multiple benchmarks and demonstrate average speedups of 3-10$\times$ with scaling properties that outperform the classical algorithms.

\end{abstract}


\section{Introduction}

% \todo[inline, color=green]{First paragraph (topic paragraph): talk about these: 1-say ISTA, FISTA and Proximal are important because they are used to solve *** in ** big data applications. 2- Since they are iterative their performance is often limited by per iteration communication. 3- Since cost of communication is larger arithmatic this communication which is tied to the formulation reduce the scalabilty and performance of the algorithms on data center and super computing architectures. Merge all the following paragraphs (until the next green box to write what I said above. Move anything extra to the background if you find necessary.}

Mathematical optimization is one of the main pillars of machine learning, where parameter values are computed based on observed data.
Applications in many big data and large-scale scientific computing domains need the solution to convex optimization problems \cite{bottou2016optimization}.
The performance of these optimization problems is often dominated by communication which is closely tied to the formulation of the algorithms.
A popular approach to estimate parameters in convex optimization problems is solving a regularized least square problem that can solve many ill-conditioned systems. 
%Regularized least square problems are often solved using the class of \textit{proximal methods}, called \textit{iterative shrinkage-thresholding algorithms (ISTA)}, where each iteration computes a matrix-vector multiply involving $X$ and $X^T$, where $X\in \mathbb{R}^{d\times n}$ is followed by a shrinkage/soft-thresholding step\cite{daubechies2004iterative,vonesch2007fast,bredies2008linear}.
L1-regularized least square problems are often solved using the class of \textit{proximal methods}, called \textit{iterative shrinkage-thresholding algorithms (ISTA)} \cite{daubechies2004iterative}. %,vonesch2007fast,bredies2008linear}.
% \todo{MG:iterative shrinkage is only for the l1 regularized pbms}
The advantage of ISTA is in its simplicity.
However, ISTA has also been recognized as a slow method in terms of convergence rate \cite{beck2009fast}. It is well known that for large-scale problems first order
methods are often the only practical option. \textit{Fast iterative shrinkage-thresholding algorithm
(FISTA)} extends ISTA while preserving the computational simplicity of ISTA and has a global rate of convergence that is significantly better, both theoretically and practically \cite{beck2009fast}. Another class of widely used proximal methods are \textit{proximal Newton-type methods} (PNM) that use second order information in their proximal mapping which have a superlinear rate of convergence \cite{lee2012proximal}. Stochastic formulations of FISTA and PNM, abbreviated as SFISTA and SPNM, are often more efficient for processing large data \cite{bottou2016optimization} and thus are used as the base algorithms in this work. These methods are iterative by nature, therefore need to communicate data in each iteration. Since communication is more expensive than arithmetic, the scalability and performance of these algorithms are bound by communication costs on distributed architectures.

The performance of an algorithm on modern computing architectures depends on the costs of arithmetic % (floating point operations (flops)) 
and data communication between different levels of memory and processors over a network. The communication cost of an algorithm is computed by adding bandwidth costs, the time required to move words, and latency costs, the time to communicate messages between processors. On modern computer architectures the cost of communication is often orders of magnitude larger than floating point operation costs and this gap is increasing \cite{dongarra2014applied}. Classical formulations of optimization methods are inherently communication-bound. Therefore, to achieve high-performance on modern distributed platforms, optimization  methods have to be reformulated to minimize communication and data movement rather than arithmetic complexity \cite{ballard2014communication}.

% Many scientific computing applications require to be executed on large scales which causes costly communication and increases running time of these application in many cases. As the size of the problem increases, the need to use resources more efficiently becomes more important.
% .Recently, some direct and iterative linear algebra algorithm have been modified to avoid communication, showing great improvement and speedups over existing libraries\cite{ballard2014communication,hoemmen2010communication,williams2014s,baboulin2012class}.

% \todo[inline, color=green]{Third Paragraph (remember when you talk about other work in your intro it has to be short and you should add another sentence right after that to say how your approach is different and solves the problem with the previous work): Here we say that most optimization problems are iterative and communicate per iteration. Since they operate on large data and because communication costs are large in distributed platforms their performance is bound by communication.  We can say their behavior can be compared to Krylov solvers used by the linear algebra community and say CA methods for them have succeeded in reducing their communication costs. But we should mention that unlike Krylov methods in scientific computing these methods use randomization and sampling to distributed data and to communicatin (I might be wrong here to its up to you what to write here).  }

Recent work has developed algorithms to reduce communication in a number of optimization and machine learning methods. Communication efficient optimization libraries such as \cite{smith2016cocoa, recht2011hogwild, zhou2017convergence} attempt to reduce communication, though they may change the convergence behavior of the algorithm. For example, CoCoA \cite{smith2016cocoa} uses a local solver on each machine and shares information between the solvers with highly flexible communication schemes. HOGWILD! \cite{recht2011hogwild} implements stochastic gradient descent (SGD) without locking which achieves a nearly optimal convergence rate (compared to its serial counterpart) only if the optimization problem is sparse.
K-AVG \cite{zhou2017convergence} modifies SGD by communicating every \textit{k} iterations. This method changes the convergence behavior of SGD by arguing that frequent synchronization does not always lead to faster convergence.
 You \textit{et. al.} \cite{you2015svm} present a partitioning algorithm to efficiently divide the training set among processors for support vector machines (SVMs). Their approach works well for SVMs but does not extend to optimization problems in general.
 P-packSVM \cite{ppacksvm} uses a similar approach to ours to derive an SGD-based SVM algorithm which communicates every {k} iterations. Our work extends this technique to proximal least-squares problems solved by FISTA and Newton-type methods. 
%Communication efficient optimization libraries have been popular recently. These frameworks attempt to reduce the communication, though they may change the convergence behavior of the algorithm. For example, CoCoA\cite{smith2016cocoa} uses local solver on each machine and share information between them in a highly flexible communication scheme. HOGWILD!\cite{recht2011hogwild} implemented stochastic gradient descent(SGD) without any locking which achieves a nearly optimal convergence rate if the optimization problem is sparse.
%K-AVG \cite{zhou2017convergence} modifies SGD through communicating every k iterations. This method changes the convergence behavior of the SGD while indicating that more frequent synchronization does not always result in faster convergence.

Standard optimization algorithms are iterative and compute an update direction per iteration by communicating data. %Since communication cost dominates the arithmetic cost of these algorithms in large scale machine learning problems, their performance is bound by communication.
%We reformulate FISTA and SPNM with structured randomization to communicate data only every \textit{``k''} iterations.
 Krylov iterative solvers, frequently used  to solve linear systems of equations, follow a similar pattern. Efforts to reformulate standard formulations of Krylov solvers by scientific computing practitioners has lead the developed of \textit{k}-step Krylov solvers \cite{chronopoulos1989efficient, CHRONOPOULOS1996623}. %Among iterative methods, the Krylov s-step methods are one of the well-known communication avoiding algorithms developed for solving linear systems. 
 These algorithms compute \textit{k} basis vectors at-once by unrolling \textit{k} iterations of the standard algorithm.% benefit from  locality by unrolling the iteration loop for s-step and compute s basis vector and perform the SpMVs operations by breaking the dependency between these operations. 
 With careful partitioning of the original data matrix  \cite{ carson2015communication,carson2013avoiding}, the matrix powers kernel in the reformulated Krylov methods can be optimized to reduce communication costs by  \textit{O(k)} compared to the classical algorithm. \textit{k}-step methods are powerful as they are arithmetically the same as the standard methods and thus often preserve convergence properties.  However, data partitioning costs can be very high in the communication-avoiding implementations of \textit{k}-step Krylov methods and matrices from many optimization problems are not good candidates for such partitioning. Devarakonda \textit{et. al.} extend \textit{k}-step methods to the block coordinate decent (BCD) methods \cite{devarakonda2016avoiding}. Their implementation reduces latency costs by a factor \textit{k} while increasing the bandwidth and floating point operations (flops) costs.% which most likely will be the primary bottleneck.
 
% These optimizations enables the algorithm to use BLAS-3 operations instead of BLAS-2 and BLAS-1 which attains higher peak performance.
%Numerous s-step Krylov methods have been designed to exploit parallelism including s-step conjugate gradients method \cite{vanrosendale1983minimizing}, s-step methods for preconditioned symmetric linear systems \cite{chronopoulos1989efficient} and s-step methods for unsymmetric linear systems \cite{CHRONOPOULOS1996623}.
%One drawback of these algorithms is the communication needed to compute s Krylov basis vectors. Proposing matrix power kernel optimization\cite{hoemmen2010communication,mohiyuddin2012tuning} reduced the communication for Krylov basis vector computation which resulted in communication avoiding Krylov subspace methods\cite{hoemmen2010communication,carson2015communication,carson2013avoiding}.
In this work we introduce \textit{k}-step formulations for the stochastic FISTA (SFISTA) and stochastic proximal Newton-type method (SPNM) algorithms. Unlike the \textit{k}-step formulations of Krylov methods which avoid communication by partitioning data---such an approach does not work for large matrices in optimization problems---we use randomized sampling to reformulate the classical algorithms and to reduce communication. Randomization enables us to control the communication and computation cost in the reformulated algorithm.
Our approach reduces the latency cost without significantly increasing bandwidth costs or flops.  

% (i.e. processors are only allowed to modify a part of the solution)
%  These methods tackle the optimization problem in a different way which changes the convergence and stability behavior of the classical algorithms.
%  This algorithm computes the Hessian and residual for s iterations and send it to all processors, while they solve a linear system redundantly.
% \todo[inline, color=red]{ SS: This paragraph talks about the effect of randomization. Where is the best place for it?}

\begin{figure}[t]
    \centering
    \includegraphics[width=0.7\linewidth]{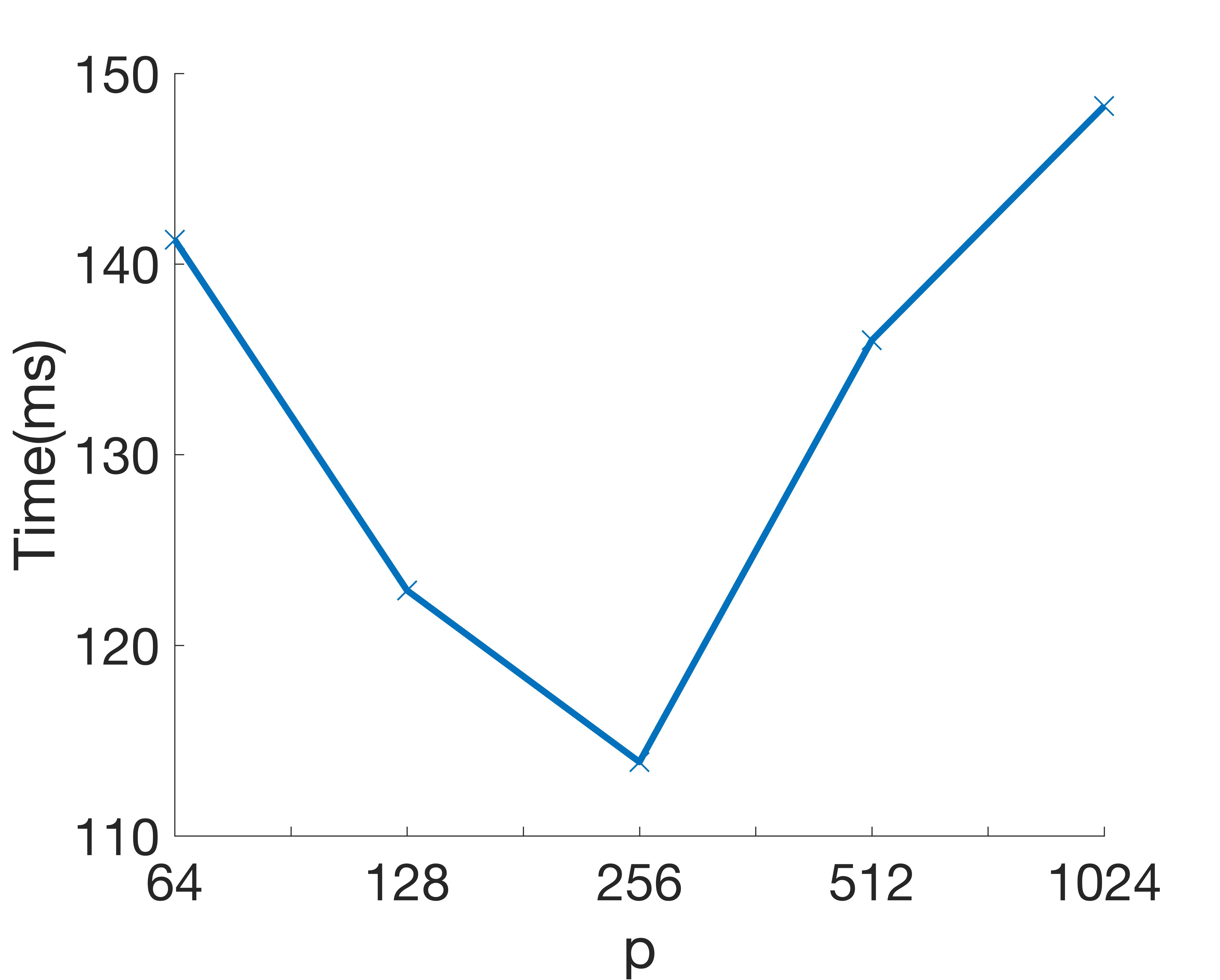}
    \caption{Execution time of SFISTA for the \textit{covtype} dataset. }
    \label{fig:nonca-cov}
    \vspace{-0.13in}
\end{figure}

We introduce communication-avoiding implementations of SFISTA (CA-SFISTA) and SPNM (CA-SPNM) to reduce communication by reducing latency costs. Figure \ref{fig:nonca-cov} shows the execution time of  SFISTA for the \textit{covtype} dataset. SFISTA shows poor scaling properties when the number of processors increase and demonstrates no performance gains on 64 processors vs. one processor. CA-SFISTA can solve the regularized least square problem by iteratively updating the optimization variable every \textit{k} iterations and reduces the latency cost by a factor of \textit{k}. CA-SFISTA and CA-SPNM outperform the classical algorithms without changing the convergence behavior. We control randomization  by sampling from the distributed data between processors. Random sampling enables us to  create sub-samples of data to improve locality and arithmetic costs in each iteration. %  benefits us in two different ways. First, sampling a mini batch from data points helps us to do more computation per iteration which could be easily tuned to get more speedups. 
Unrolling of iterations to reformulate the standard algorithms is also made possible through random sampling.

% \todo[inline,color=yellow]{AD: abalone is a tiny dataset, do you have data for SUSY or Covtype. I have data for them. Should I replace it with another one?}

The following summarizes the contributions of the work:
\begin{itemize}
  \item We introduce CA-SFISTA and CA-SPNM that provably reduce the latency cost of two well-known optimization methods used to solve L1-regularized least square problems by a factor of $O(k)$ without changing the convergence behavior of the original algorithms.  
    \item The communication-avoiding algorithms achieve upto 10X speedup on the target supercomputers.
  \item We develop a randomized sampling strategy that creates samples of data for each processor to facilitate the derivation and implementations of CA-SFISTA and CA-SPNM. 
 % \todo[inline, color=red]{I think we should remove the next one}
%  \item While communication-avoiding formulations of Krylov solvers suffer from a costly and sometimes unsuccessful matrix partitioning phase, we leverages randomized sampling---already used in standard stochastic optimization methods---for data distribution and to avoid communication.

  \item Computation and communication costs of the classical algorithms as well as the proposed communication-avoiding algorithms are presented. Our analysis shows that CA-SPNM and CA-SFISTA  reduce communication costs without changing the floating points operation or bandwidth costs. 
\end{itemize}

% \subsection{Organization}
% "To be filled..."
\section{Background}

% \todo[inline, color=green]{Each section needs a topic paragraph.} 
This section introduces the L1-regularized least squares problems and algorithmic approaches for their solution. Details of the communication model used throughout the paper is also provided. % in the next section and then we move to the communication model we use for measuring the performance of our algorithms  terms of the communication and computation costs. %\ref{sec:comm-model}.

% \todo{MG:This paragraph needs further clarification.}
% \todo[inline,color=red]{SS: Should I explain lasso more though I've cited the paper?}

\subsection{Regularized Least Squares Problem}
\label{sec:lasso}

% \todo[inline,color=green]{Why dont you start with the generalized optimization problem form (what you have in IV and then say how Li regularized problem is its special case. Also howare FISTA, proximal, and ISTA connected to this.}
Consider the following composite additive cost optimization problem of the form 
\begin{equation}
\label{eq:minimization}
\operatornamewithlimits{min}\limits_{w\in \mathbb{R}^d} \;{F(w)\equiv f(w)+g(w)}
\end{equation}
where $g:\mathbb{R}^d \to \mathbb{R}$ is a continuous, convex, possibly nonsmooth function and~$f:\mathbb{R}^d \to \mathbb{R}$~is a smooth and convex function, with a Lipschitz-continuous gradient. Let $L(f)$ be the Lipschitz constant of the gradient of $f$. This form can represent a large class of regression problems based on different choices of $f$ and $g$. In particular, the \emph{L1-regularized least squares problem}, a.k.a. the least absolute shrinkage and selection operator (LASSO) problem, is a special case where $$f(w)=\frac{1}{2n}\|X^Tw-y\|_{2}^2, \quad g(w)=\lambda \|w\|_1,$$
when $X\in \mathbb{R}^{d\times n}$ is the input data matrix, where rows are the features and columns are the samples, $y\in \mathbb{R}^n$ holds the labels (or observations), $w \in \mathbb{R}^d$ is the optimization variable, and $\lambda \in \mathbb{R}$ is the regularization (penalty) parameter. The parameter $\lambda$ controls the sparsity of the solution, since increasing its value magnifies the effect of the second term $g(w)$, which itself is minimized at $w=0$. In this case, the gradient of $f(w)$ is given by:\\
\begin{equation}
\label{eq:gradient}
\nabla f(w)=\frac{1}{n}(XX^Tw-Xy).\
\end{equation}

LASSO problems appear frequently in machine learning applications \cite{tibshirani1996regression} including learning directed and undirected graphical models \cite{schmidt2007learning}, online dictionary learning \cite{mairal2009online}, elastic-net regularized problems \cite{smith2015l1}, and feature
selection in classification problems and data analysis \cite{rendle2016robust,koh2007interior}. 
%LASSO attempts to find the parameter $\hat{w}$ that minimize and objective function which has two parts: a quadratic penalty $f$ and the L1-regularization term $g$. The optimal parameters are obtained by solving the following optimization problem.
Let 
\begin{equation}
\hat{w}=\operatornamewithlimits{argmin}\limits_{w}  \frac{1}{2n}\|X^Tw-y\|^2 +\lambda \|w\|_1
\end{equation}
be the optimal solution to the LASSO problem. The LASSO objective contains two parts, a quadratic penalty $f$ and the L1-regularization term $g$.  Therefore, LASSO has regression characteristics by minimizing a least squares error and has subset selection features by shrinking the optimization variable and setting parts of it to zero.

% \todo[inline,color=green]{Add a subsection and talk about FISTA, ISTA, and proximal NM in a high level. As the reader I am still confused about the connection between regularized least square, FISTA, ISTA, and proximal methods. You might have to move text from upcoming sections to here to clarify this.}
\subsection{Solving LASSO}
Among the most efficient methods for solving convex optimization problems of the form  \eqref{eq:minimization} are first order methods that use a proximal step in order to handle the possibly non-smooth part $g$. In particular, for the LASSO problem, the class of iterative soft thresholding algorithms, recently improved to  FISTA \cite{beck2009fast}, have become very popular due to their excellent performance. Each iteration of FISTA, involves computing a gradient followed by a soft thresholding operator. More recently, inspired by first order methods, proximal Newton methods have been introduced that use second order information (hessian) in their proximal part. These methods are globally convergent and could achieve a superlinear rate of convergence in the neighborhood of the optimal solution \cite{lee2012proximal}. The fact that FISTA and proximal Newton methods are key algorithms for solving composite convex optimization problems motivates us to introduce communication-avoiding variants of these methods and discuss their computation and communication complexity.

\subsection{Communication Model}
\label{sec:comm-model}
The cost of an algorithm includes arithmetic and computation. Traditionally, algorithms have been analyzed with floating point operation costs. However, communication costs are a crucial part in analyzing algorithms in large-scale simulations \cite{fuller2011computing}. The cost of floating point operations and communication, including bandwidth and latency, can be combined to obtain the following model: 
\begin{equation}
T = \gamma F+\alpha L+\beta W
\end{equation}
where $T$ is the overall execution time approximated by a linear function of F, L, and W, which are total number of floating point operations, the number of messages sent, and the number of words moved respectively. Also, $\gamma$, $\alpha$, and $\beta$ are machine-specific parameters that show the cost of one floating point operation, the cost of sending a message, and the cost of moving a word. 
Among different communication models, the LogP \cite{culler1993logp} and LogGP \cite{alexandrov1995loggp} models are often used to develop communication models for parallel architectures. The communication model used in this paper is a simplified model known as $\alpha-\beta$ which uses $\gamma$, $\alpha$, and $\beta$ and shows the effect of communication and computation on the algorithm cost.

\section{Classical Stochastic algorithms}
\label{sec:classical-alg}
Even though there are some advantages to using batch optimization methods, multiple intuitive, practical, and theoretical reasons exist for using stochastic optimization methods. Stochastic methods can use information more efficiently than batch methods by choosing data randomly from the data matrix. In particular, for large-scale machine learning problems, training sets involve a good deal of (approximate) redundant data which makes batch methods inefficient in-practice. Theoretical analysis also favors stochastic methods for many big data scenarios \cite{bottou2016optimization}.
This section explains the SFISTA and SPNM algorithms and analyzes their 
%a stochastic variant of FISTA (SFISTA) and stochastic proximal Newton method (SPNM)
computation and communication costs. % are analyzed for both algorithms.% and the communication bottleneck of SFISTA and SPNM.
The associated costs are derived under the assumption that columns of $X$ are distributed in a way that each processor has roughly the same number of non-zeros. The vector $y$ is distributed among processors while vectors with dimension $d$ such as $v$ and $w$ are replicated on all processors. Finally, we assume that $n \gg d$ which means we are dealing with many observations in the application.

\subsection{Stochastic Fast Iterative Shrinkage-Thresholding Algorithm (SFISTA)}
% 		\todo[color=yellow, inline]{AD: $\Delta w_{j-1}$ is never computed in Alg. 1.}
%         \todo[inline,color=red]{I've explained $\Delta w$ right before subsection B. Do I need to put it in algorithm I?}
%         \todo[inline,color=green]{I think AD wants $\Delta w_{j-1}$ or $w$ to be declared in the algorithms ?}

% \todo[inline,color=green]{You never explained why the gradient of f is needed. You need to tell the reader that Table I shows FISTA, and demonstrates that a gradient has to be computed per iteration. But since the gradient is expensive to compute a sampled form is often used. Explicitly mention that sampling is used by the ML community to reduce computation costs and also provides acceptable convergence guarantees.}

% FISTA algorithm solves the following generalized optimization problem:
% \begin{equation}
% \label{eq:minimization}
% \operatornamewithlimits{min}\limits_{w\in \mathbb{R}^d} \;{F(w)\equiv f(w)+g(w)}
% \end{equation}
% where $f(w)$ is a smooth convex function, continuously differentiable with Lipschitz continuous gradient L(f) and $g(w)$ is a continuous convex function which is possibly nonsmooth. The L1-regularized least square problem is a special case by substituting $f(w)=\frac{1}{2n}\|X^Tw-y\|_{2}^2$ and $g(w)=\lambda \|w\|_1$. The gradient of $f(w)$ is:\\
% \begin{equation}
% \label{eq:gradient}
% \nabla f(w)=\frac{1}{n}(XX^Tw-Xy)\
% \end{equation}
Computing the gradient in (\ref{eq:gradient}) is very expensive since the data matrix $X$ is being multiplied by its transpose. The cost of this operation can be reduced by sampling from the original matrix, leading to a randomized variant of FISTA \cite{xiao2014proximal,nitanda2014stochastic}. If $b$ percent of columns of $X$ are sampled, the new gradient vector is obtained by:
\begin{equation}
\label{eq:gradient-upd}
\nabla f(w)=\frac{1}{m}(X I_j I_j^T X^Tw-XI_j I_j^T y)\
\end{equation}
where $m=\floor{bn}$ and $I_j$ is a random matrix containing one non-zero per column representing samples used for computing the gradient.
 Therefore the generalized gradient update is:\\
 \begin{equation}
 w_{j+1}=S_{\lambda t}(w_j-t_j\nabla f(w_j))
 \end{equation}
 where $S_{\lambda}$ is the soft thresholding operator defined as:
 \begin{equation}
 S_{\lambda}(w)]_i=\begin{cases}
 
 w_i-\lambda  & \text{if}\ w_i>\lambda\\
 0 & \text{if}\ -\lambda \leq w_i\leq \lambda\\
 w_i+\lambda & \text{if}\ w_i<-\lambda
 
 \end{cases}
 \end{equation}

\noindent
where $[.]_i$ represents the \textit{i}-th element of a vector. FISTA accelerates the rate of generalized gradient by modifying the update rule as follows:
 \begin{equation}
  w_{j+1}=S_{\lambda t}(v_j-t_j\nabla f(w_j))
 \end{equation}
 where $v_j$ is the auxiliary variable defined as:
 \begin{equation}
  v_j=w_j+\frac{j-2}{j}\Delta w_j\
 \end{equation}
 and $\Delta w_j=w_j-w_{j-1}$.
Algorithm \ref{tab:fista-alg} shows the SFISTA algorithm for solving the LASSO problem.\\

\subsection{Stochastic Proximal Newton Methods (SPNM)}
% \todo[inline,color=green]{Proximal has to be a new subsection. Also first start that subsection by saying why FITA is not always good and why we need proximal too. Be more clear in connecting proximal to FISTA. If the reader has learnt Table I now they are interested to see in which lines and why does proximal differ. Say what is H.}
FISTA uses the first order information of $f$ (gradient) to update the optimization variable. However, proximal Newton methods solve the same problem with second order information (Hessian), or an approximation of it, to compute the smooth segment of the composite function (\ref{eq:minimization}). Proximal Newton methods achieve a superlinear convergence rate in the vicinity of the optimal solution \cite{lee2012proximal}. As discussed in section \ref{sec:lasso}, since LASSO starts with an initial condition at $w=0$ and because the optimal solution is typically very sparse, the 
%the optimal solution usually is sparse and lots of its element are zero, one can expect that the 
sequence of $w_k$ will be close enough to the optimal solution for different values of \textit{k}. Therefore, very often proximal Newton methods achieve a superlinear convergence rate and converge faster than FISTA. % to the optimal solution. 
Proximal Newton methods solve (\ref{eq:minimization}) with quadratic approximation to the smooth function $f$ in each iteration \cite{lee2012proximal} as follows:
\begin{equation}
\label{eq:proximal}
\begin{split}
w_{j+1}&=\operatornamewithlimits{argmin}\limits_{y} \nabla f(w_j)^T(y-w_j)\\
&+\frac{1}{2}(y-w_j)^T H_j (y-w_j)+g(y)
\end{split}
\end{equation}
where $H_j$ is the approximation of the Hessian $f$ at iteration $j$.  Since a closed form solution of \eqref{eq:proximal} does not exist, a first order algorithm is often used to solve \eqref{eq:proximal} and to update the optimization variable.  Algorithm \ref{tab:spnm-alg} shows the stochastic proximal Newton method for solving LASSO. As demonstrated, the SPNM algorithm  takes a block of data, approximates the Hessian based on sampled columns, and uses a first order solver to solve \eqref{eq:proximal} in lines 7-10 with $Q$ updates on the optimization variable. Thus, SPNM can be seen as a first order solver operating on blocks of data.

% \begin{table}
% 	\caption{FISTA Algorithm }
%     \centering
% 	\label{tab:fista-alg}
% 	\begin{tabular}{|ll|p{0.2cm}|p{2cm}|}
% 		\hline
        
%         \multicolumn{2}{|c|}{Algorithm 1 FISTA Algorithm}\\
% 		\hline
		
% 		1: & \textbf{Input:} $X \in \mathbb{R}^{d \times n}$, $y \in \mathbb{R}^{n}$, $w_0 \in \mathbb{R}^d$, K$>$1.  \\
% 		2: & \textbf{for} $k=0,1,...,K$ do\\
%        3:& \quad  $\nabla f(w_{k})=\frac{1}{n}XX^Tw_{k-1}-\frac{1}{n}XX^Ty$\\

%        4:& \quad $v_{k}=w_{k-1}+\frac{k-2}{k}(\Delta w_{k-1})$\\
%        5:&\quad  $w_{k}=S_{\lambda t_k}(v_{k}-t_k \nabla g(w_{k}))$\\
%        6:& \textbf{output} $w_K$\\
% 		\hline
		
% 	\end{tabular}
% \end{table}
\captionsetup[table]{name=Algorithm}
\begin{table}
    \caption{The SFISTA algorithm }
    \centering
	\label{tab:fista-alg}
	\begin{tabular}{|ll|p{0.2cm}|p{2cm}|}
% 		\hline
        
%         \multicolumn{2}{|c|}{Algorithm 1 FISTA Algorithm}\\
		\hline
		
		1: & \textbf{Input:} $X \in \mathbb{R}^{d \times n}$, $y \in \mathbb{R}^{n}$, $w_0 \in \mathbb{R}^d$, K$>$1 and $b\in \left(0,1 \right]$.  \\
        2:& set $m=\floor{bn}$ \\
		3: & \textbf{for} $j=0,1,...,T$ do\\
        4:& \quad Generate $I_{j}=[e_{i_1},e_{i_2},...,e_{i_m}]\in \mathbb{R}^{n\times m}$ where\\
	&\quad  $\{i_{h}\in[d]|h=1,...,m\}$ is chosen uniformly at random \\
    
       5:& \quad  $\nabla f(w_{j})=\frac{1}{m}XI_j I_j^T X^T w_{j-1}-\frac{1}{m}X I_j I_j^T y$\\

       6:& \quad $v_{j}=w_{j-1}+\frac{j-2}{j}(\Delta w_{j-1})$\\
       7:&\quad  $w_{j}=S_{\lambda t_j}(v_{j}-t_j \nabla f(w_{j}))$\\
       8:& \textbf{output} $w_T$\\
		\hline
	\end{tabular}
\end{table}
% \begin{table}
% 	\caption{SPNM Algorithm }
%     \centering
% 	\label{tab:spnm-alg}
% 	\begin{tabular}{|ll|p{0.2cm}|p{2cm}|}
% 		\hline
%         \multicolumn{2}{|c|}{Algorithm 2 SPNM Algorithm}\\
% 		\hline
% 		%\textbf{Hamrle2} &circuit simulation problem & 1.1 & 1.04 & \\
% 		%\hline
% 		1: & \textbf{Input:} $X \in \mathbb{R}^{d \times n}$, $y \in \mathbb{R}^{n}$, $w_0 \in \mathbb{R}^d$, K$>$1.  \\
		
%        2:& \quad  $\nabla f(w_{k})=\frac{1}{n}XX^Tw_{k-1}-\frac{1}{b}XX^Ty$\\

%        3:& \quad \quad $z_{0}=w_{k-1}$\\
%        4:&\quad \quad \textbf{for} $q=1,...,Q$\\
%        5:&\quad \quad \quad $z_{q}=S_{\lambda t_k}(z_{q-1}-t_k \nabla g(z_{q}))$\\
%        6:&\quad \quad \quad $q=q+1$\\
%        7:& \quad \quad $w_{k}=z_{Q}$\\
%        8:& \textbf{output} $w_K$\\
% 		\hline
		
% 	\end{tabular}
% \end{table}

\begin{table}
	\caption{The SPNM algorithm }
    \centering
	\label{tab:spnm-alg}
	\begin{tabular}{|ll|p{0.2cm}|p{2cm}|}
% 		\hline
%         \multicolumn{2}{|c|}{Algorithm 2 SPNM Algorithm}\\
		\hline
		%\textbf{Hamrle2} &circuit simulation problem & 1.1 & 1.04 & \\
		%\hline
		1: & \textbf{Input:} $X \in \mathbb{R}^{d \times n}$, $y \in \mathbb{R}^{n}$, $w_0 \in \mathbb{R}^d$, K$>$1 and $b\in \left(0,1 \right]$.  \\
        2:& set $m=\floor{bn}$ \\
        3: & \textbf{for} $j=0,1,...,T$ do\\
		4:& \quad Generate $I_{j}=[e_{i_1},e_{i_2},...,e_{i_m}]\in \mathbb{R}^{n\times m}$ where\\
	&\quad  $\{i_{h}\in[d]|h=1,...,m\}$ is chosen uniformly at random \\
       5:& \quad  $\nabla f(w_{j})=\frac{1}{m}XI_j I_j^T X^T w_{j-1}-\frac{1}{m}X I_j I_j^T y$\\

       6:& \quad \quad $z_{0}=w_{j-1}$\\
       7:&\quad \quad \textbf{for} $q=1,...,Q$\\
       8:&\quad \quad \quad $z_{q}=S_{\lambda t_k}(z_{q-1}-t_j \nabla f(z_{q}))$\\
       9:&\quad \quad \quad $q=q+1$\\
       10:& \quad \quad $w_{j}=z_{Q}$\\
       11:& \textbf{output} $w_T$\\
		\hline
		
	\end{tabular}
\end{table}

%  \todo[inline,color=green]{In the next paragraph you mention that you are introducing stochastic FITSA, but no where before this was explicitly explained so the reader gets confused.}
 
%  \todo[inline,color=green]{Maybe instead to trying to sell your stochastic algorithm here just stick to the standard sampled method an finish this section with its communication analysis. You can add a last paragraph to this section that mentioned the problem with the standard algorithms and that your work proposes a k-step method that also does randomization which reduces *** problem.}
 
% The reason that we are introducing stochastic FISTA is that computing the gradient vector and Hessian requires a costly operation which in many cases is not possible. Stochastic methods are the very first and popular methods to overcome this bottleneck. Therefore, we use randomization as an strategy to control the amount of computation on every processor and as a way to reduce latency at the same time.

The following theorems analyze the computation and communication costs of SFISTA and SPNM.

\textbf{Theorem 1.} $T$ iterations of SFISTA on $P$ processors over the critical path has following costs:
F = $O(\frac{Td^2bn}{P})$ flops, W = $O(Td^2log P)$  words moved, L = $O(T log P)$ messages and M = $O(\frac{dn}{P})$ words of memory.\\
\textit{Proof.} SFISTA computes the gradient in line 5 which consists of three parts. The first part multiplies $X$ by its transpose in line 5 which requires $O(\frac{d^2bn}{P})$ operations and communicates $O(d^2log P)$ words and $O(log P)$ messages. Computing the second part which involves multiplying sampled $X$ and $y$, requires $O(\frac{dbn}{P})$ operations and needs $O(dlog P)$ words with $O(log P)$ messages. These two operations dominate other costs of the algorithm. Finally, the algorithm computes the gradient and updates the optimization variable redundantly on processors. Computing the gradient (line 5) requires $O(d^2)$ operations without any communication between processors. The vector updates need $O(d)$ operations. Therefore, the total cost of SFISTA for $T$ iterations is $O(\frac{Td^2bn}{P})$ flops, $O(Td^2log P)$ words, and $O(T log P)$ messages. Each processor needs enough memory to store three parts of $\nabla f$,$v_{j}, w_{j}$ and $\frac{1}{P}$-th of $X$. Therefore, it needs $d^2+d+d+d+d+\frac{dn}{P}=O(\frac{dn}{P})$ words of memory.

% \todo[color=yellow, inline]{AD: Does bound on memory implicitly assume $d^2 < \frac{n}{P}$? How about for THM 2.}
% \todo[inline, color=red]{yes. Since we assume n is much bigger than d}

\textbf{Theorem 2.} T iterations of SPNM on $P$ processors over the critical path has the following costs:
F = $O(\frac{Td^2bn}{P}+\frac{Td^2}{\epsilon})$ flops, W = $O(Td^2log P)$  words moved, L = $O(T log P)$ messages, and M = $O(\frac{dn}{P})$ words of memory.\\
\textit{Proof.} SPNM solves the minimization problem using the Hessian. This requires $O(\frac{1}{\epsilon})$ inner iterations in order to reach an $\epsilon$-optimal solution. An analysis similar to theorem 1 can be used to prove this theorem.

{\textit{The communication bottleneck of SFISTA and SPNM:}} Despite the fact that stochastic implementations of FISTA and proximal Newton methods are more practical for large-scale machine learning applications, they do not scale well on distributed architectures (e.g. Figure \ref{fig:nonca-cov}). In each iteration of both algorithms, the gradient vector has to be communicated at line 5 with an all-reduce operation  which leads to expensive communication.

\section{Avoiding Communication in SFISTA and SPNM}
%\input{Sources/sec-algorithm}

% \todo[inline, color=green]{rename this section to complexity analysis, also add topic paragraph! } 
\label{sec:ca-alg}
% \todo[inline,color=green]{This section is our main contribution section so it has to be more clear and have some excitement to it! The topic paragraph should say We resolve the ** and ** issue by reformulating ** and proposed a novel randomization ***. We will demonstrate how our reformulation and its implementation improves communication and the algorithm scalability.}

% \todo[inline,color=green]{Do not reference previous sections. Back or forward reference in text is generally not recommended. Also reduce the topic paragraph to only one paragraph.}

% In this section , we analyze the cost for classical and communication avoiding algorithms, including computation, memory, bandwidth and latency costs. We derive the associated costs under the assumption that the columns of $X$ are distributed in a way that each processor has roughly the same amount of non-zeros. The vector $y$ is distributed among processors while vectors with dimension $d$, i.e. $v$ and $w$ are stored on all processors. We assume the computed matrices $G$ and $R$ are dense. We begin our analysis by analyzing the complexity cost of classical algorithms in section \ref{analysis:classical} and compare it to their variants in section \ref{analysis:ca}.

We reformulate the SFISTA and SPNM algorithms to take \textit{k}-steps at-once without communicating data. The proposed \textit{k}-step formulations reduce data communication and latency costs by a factor of \textit{k} without increasing bandwidth costs and significantly improve the scalability of SFISTA and SPNM on distributed platforms. 
This section presents our formulations for \textit{k}-step SFISTA and \textit{k}-step SPNM, also referred to as  communication-avoiding SFISTA (CA-SFISTA) and SPNM (CA-SPNM). We will also introduce a randomized sampling strategy that leverages randomization to generate data partitions for each processor and to make the \textit{k}-step derivations possible. % reduce the  was used in our reformulations to  we discuss the way we employ randomization and its importance in derivation of communication avoiding algorithms. 
Communication, memory, bandwidth, and latency costs for the CA algorithms are also presented.%, including computation, memory, bandwidth and latency costs and demonstrate that our reformulation enhances the scalability and avoids communication compared to the classical algorithms which enables us to get a better speedup on supercomputers. %Finally, we end this section with a discussion on implementing the algorithm on a modern super computer.

\subsection{ The CA-SFISTA and CA-SPNM Algorithms}
% \todo[inline,color=green]{Add more text and explicitly discuss the algorithm, you dont need to insert formulations but point to line number. Also add a subsection here that talks about random sampling and clearly discuss how randomization allowed you to reformulated. }

%%%%%%%%%%%%%%%%%
%%%%%%%%%%%%%%%%%
%%%%%%%%%%%%%%%%%
%%%%%%%%%%%%%%%%%
%%%%%%%%%%%%%%%%% FISTA:

% \todo[inline,color=green]{Remove references to previous sections. Its fine to point to a previous table or algorithm. Also rename Tables I to IV to Algorithms I to IV and fix all references to it.}

\textit{Communication-avoiding SFISTA:} Algorithm \ref{tab:ca-fista-alg} shows the CA-SFISTA algorithm. As discussed, the communication bottleneck is at line 5 in Algorithm \ref{tab:fista-alg}, thus for the reformulation we start by unrolling the loop in line 4 of the classical algorithm for \textit{k}-steps and will sample from the data matrix $X$ to compute the gradient of the objective function. As demonstrated in Algorithm \ref{tab:ca-fista-alg}, a random matrix is produced based on uniform distribution and is used to select columns from of $X$ and $y$ and to compute Gram matrices $G_j\in \mathbb{R}^{d\times d}$ and $R_j\in \mathbb{R}^{d}$ in line 6. Operations in line 6 are done in \textit{k} unrolled iterations because every iteration involves a different random matrix $I_{ik+j}$; without randomized sampling we could not unroll these iterations. %Iteration unrolling is possible because every iteration involves a different random matrix $I_{ik+j}$% and a different matrix-matrix multiplication. 
%These operation  could be done in \textit{k} unrolled iterations. Without randomized sampling we could not unroll these iterations.

%\todo[color=green, inline]{modify the block sentences to reflect why and how blocks of Gram matricies are chosen}

These local Gram matrices are concatenated into matrices $G\in \mathbb{R}^{d\times kd}$ and $R\in \mathbb{R}^{d \times k}$ in line 7 which are then broadcasted to all processors. This communication only occurs every \textit{k} iterations, thus, the CA-SFISTA algorithm reduces latency costs. Also, sending large amounts of data at every \textit{k} iteration improves bandwidth utilization. Processors do not need to communicate for the updates in lines 9-13 for \textit{k} iterations.  Gram matrices $G$ and $R$ consist of $k$ blocks of size $d\times d$ and  $k$ vectors of size $d$ respectively. At every iteration a block of $G$ and one column of $R$ is chosen and is used to compute the gradient in line 10. Each block of $G$ and $R$ comes from sub-sampled data and contributes to computing the gradient. The auxiliary variable is updated in line 12 and the soft thresholding operator updates $w$. To conclude, the CA-SFISTA  algorithm only communicates data every \textit{k} iterations in line 7 which reduces the number of messages communicated between processors by \textit{O(k)}.

% \todo[color=green, inline]{line 7 in Algorithms III and IV are aligned with the for loop in line 3 which means line 3 for loops finished at line 7. I think this line should ab aligned with line 4 for loop instead..}

\begin{table}
\centering
	\caption{The CA-SFISTA Algorithm }
	\label{tab:ca-fista-alg}
	\begin{tabular}{|ll|p{0.2cm}|p{2cm}|}
% 		\hline
%         \multicolumn{2}{|c|}{Algorithm 3 Communication-Avoiding 
%         FISTA (CA-FISTA) Algorithm}\\
		\hline
		%\textbf{Hamrle2} &circuit simulation problem & 1.1 & 1.04 & \\
		%\hline
		1: & \textbf{Input:} $X \in \mathbb{R}^{d \times n}$, $y \in \mathbb{R}^{n}$, $w_0 \in \mathbb{R}^d$, K$>$1, $b \in \mathbb{Z}_{+}$ s.t $b \leq n$  \\
        2:& set $m=\floor{bn}$ \\
		3: & \textbf{for} $i=0,1,...,\frac{T}{s}$ do\\
		
		4: & \quad \textbf{for} $j=1,...,k$ do  \\
		5:  & \quad \quad  Generate $I_{ik+j}=[e_{i_1},e_{i_2},...,e_{i_m}]\in \mathbb{R}^{n\times m}$ where\\
	&\quad \quad  $\{i_{h}\in[d]|h=1,...,m\}$ is chosen uniformly at random \\
       6: & \quad \quad $G_j=\frac{1}{m}XI_{ik+j}I_{ik+j}^TX^T, R_j=\frac{1}{m}XI_{ik+j}I_{ik+j}^Ty $ \\
       7: &  \quad set $G=[G_1|G_2|...|G_k]$ and $R=[R_1|R_2|...|R_k]$ \\
          & \quad and send them to all processors.\\
       8: & \quad \textbf{for} $j=1,...,k$ do \\
       9: & \quad \quad $H_{ik+j}$ are $d\times d$ blocks of G\\
       10: & \quad \quad $\nabla f(w_{ik+j})=H_{ik+j}w_{ik+j-1}-R_{ik+j}$\\
       11:& \quad \quad $w_{ik+j}=\operatornamewithlimits{argmin}\limits_{y} \nabla f(w_{ik+j})^T (y-w_{ik+j-1})$\\
          &\quad \quad \quad \quad \quad \quad $+\frac{1}{2}(y-w_{ik+j-1})^T (y-w_{ik+j-1})+h(y)$\\
          & \quad \quad solve the optimization using FISTA:\\
       12:& \quad \quad $v_{ik+j}=w_{ik+j-1}+\frac{ik+j-2}{ik+j}(w_{ik+j-1}-w_{ik+j-2})$\\
       13:&\quad \quad $w_{ik+j}=S_{\lambda t_{ik+j}}(v_{ik+j}-t_{ik+j} \nabla g(w_{ik+j}))$\\
       14:& \textbf{output} $w_T$\\
		\hline
		
	\end{tabular}
\end{table}

%%%%%%%%%%%%%%%%%
%%%%%%%%%%%%%%%%%
%%%%%%%%%%%%%%%%%
%%%%%%%%%%%%%%%%%
%%%%%%%%%%%%%%%%% SPNM:
\textit{Communication-avoiding SPNM:} Similar to the CA-SFISTA formulation, CA-SPNM in Algorithm \ref{tab:ca-spnm-alg} is formulated by unrolling iterations in line 4 of the classical SPNM algorithm. Lines 1 through 10 in Algorithm \ref{tab:ca-spnm-alg} follow the same analysis as CA-SFISTA.  CA-SPNM solves the inner subproblem inexactly in lines 13-16. It uses a first order method without any communication to get an $\epsilon$-optimal solution for the inner problem. The same blocks of $G$ and $R$ are used in the subproblem until a solution is achieved after $Q$ iterations. The value of $w$ from the previous iteration is used as a \textit{warm start} initialization in line 13 to improve the convergence rate of the inner iterations and the overall algorithm. The CA-SPNM  algorithm  only communicates the Gram matrices at line 7 and thus the total number of messages communicated is reduced by a factor of \textit{k}.

In conclusion, derivations of CA-SFISTA and CA-SPNM start from a randomized variant of the classical algorithm, enabling us to unroll the iterations while maintaining the exact arithmetic of the classical algorithms. %Then computing Gram matrices based on the different randomized matrices helped us to form $G$ and $R$ before updating variables in both algorithms resulting in a k-step method to reduce communication cost of classical algorithms. In next section, we deep into randomization and show that \textit{k}-step methods are possible to derive when we employ this randomization to benefit us in different aspects.

\begin{table}
\centering
	\caption{The CA-SPNM Algorithm }
	\label{tab:ca-spnm-alg}
	\begin{tabular}{|ll|p{0.2cm}|p{2cm}|}
% 		\hline
%         \multicolumn{2}{|c|}{Algorithm 4 Communication-Avoiding 
%         SPNM (CA-SPNM) Algorithm}\\
		\hline
		%\textbf{Hamrle2} &circuit simulation problem & 1.1 & 1.04 & \\
		%\hline
		1: & \textbf{Input:} $X \in \mathbb{R}^{d \times n}$, $y \in \mathbb{R}^{n}$, $w_0 \in \mathbb{R}^d$, K$>$1, $b \in \mathbb{Z}_{+}$ s.t $b \leq n$  \\
        2:& set $m=\floor{bn}$ \\
		3: & \textbf{for} $i=0,1,...,\frac{T}{s}$ do\\
		
		4: & \quad \textbf{for} $j=1,...,k$ do  \\
		5:  & \quad \quad  Generate $I_{ik+j}=[e_{i_1},e_{i_2},...,e_{i_m}]\in \mathbb{R}^{n\times m}$ where\\
	&\quad \quad  $\{i_{h}\in[d]|h=1,...,m\}$ is chosen uniformly at random \\
       6: & \quad \quad $G_j=\frac{1}{m}XI_{ik+j}I_{ik+j}^TX^T, R_j=\frac{1}{m}XI_{ik+j}I_{ik+j}^Ty $ \\
       7: & \quad set $G=[G_1|G_2|...|G_s]$ and $R=[R_1|R_2|...|R_s]$\\
          & \quad and send them to all processors\\
       8: & \quad \textbf{for} $j=1,...,k$ do \\
       9: & \quad \quad $H_{sk+j}$ are $d\times d$ block of G\\
       10: & \quad \quad $\nabla f(w_{ik+j})=H_{ik+j}w_{ik+j-1}-R_{ik+j}$\\
       11:& \quad \quad $w_{ik+j}=\operatornamewithlimits{argmin}\limits_{y} \nabla f(w_{ik+j})^T (y-w_{ik+j-1})$\\
          &\quad \quad \quad \quad \quad$+\frac{1}{2}(y-w_{ik+j-1})^T H_{ik+j}(y-w_{ik+j-1})+h(y)$\\
       12:&\quad \quad *solve this minimization problem using a first order method\\
       13:& \quad \quad $z_{0}=w_{ik+j-1}$\\
       14:&\quad \quad \textbf{for} $q=1,...,Q$\\
       15:&\quad \quad \quad $z_{q}=S_{\lambda t_{ik+j}}(z_{q-1}-t_{ik+j} \nabla f(z_{q}))$\\
       16:&\quad \quad \quad $q=q+1$\\
       17:& \quad \quad $w_{ik+j}=z_{Q}$\\
       18:& \textbf{output} $w_T$\\
		\hline
	
	\end{tabular}
\end{table}
\captionsetup[table]{name=Table}
\setcounter{table}{0}

\subsection{Using Randomized Sampling to Avoid Communication}
\label{sec:rand-benefit}

%  \todo[inline,color=green]{So the first and the second paragraph here say the same thing. Can you merge them? Try not to change the writing in the first paragraph since I edited that. We can remove paragraph 2 if its saying the same thing. Paragraph 3 is already edited by me so leave it as is.}
 
%  \todo[color=green, inline]{Also I recommend moving your Table I on the implementation to this section and maybe modifying the text in this section (and even the Table I algorithm) to show where and how randomization is done in the implementation. Maybe you can start the second paragraph by saying ``Table I shows a pseudo implementation of  CA-FISTA (CA-SPNM follows a similar pattern). While the data matrix $X$ distributed amongst different processors in line 3, randomize sampling is used in line 6 to ****." I would also change the line number reference in the third paragraph of this section to point to Table I.  }
 
%  \todo[color=green, inline]{Table I is not a full pseudo code, so lets call it the implementation of CA-SFISTA. Also describe how you create blocks in line 11 in the table. Finlay call this an algorithm not a table. }
 
We leverage randomized sampling in the stochastic variants of FISTA and proximal Newton methods to derive the communication-avoiding formulations. %The communication-avoiding SFISTA and SPNM reformulate the stochastic variants of FISTA and proximal Newton methods to avoid communication.   
%The k-step methods CA-SFISTA and CA-SPNM are derived from a randomized variant of classical algorithms. 
With randomized sampling, CA-SFISTA and CA-SPNM generate independent samples at each iteration. These randomly selected samples contribute to computing gradient and Hessian matrices and as a result allow us to unroll \textit{k} iterations of the classical algorithms to avoid communication. We create blocks of the Gram matrices $G$ and $R$ by randomly selecting \textit{k} different subset of the columns by each processor. Performing one all-reduce operation every \textit{k} iterations on these Gram matrices is far less expensive than doing an all-reduce operation at every iteration, which enables us to avoid communication.

% We exploit this sub-sampling nature of stochastic algorithms to develop an algorithm that reduces latency cost, preserve bandwidth and flops cost and uses bandwidth better compared to classical algorithms. Stochastic FISTA and SPNM help us to select bunch of columns at each iteration, compute gradient using these columns and perform an update on the variables. 
% These Gram matrices consist of the information for \textit{k} iterations.
Algorithm \ref{tab:pseudo-fista} shows a pseudo implementation of CA-SFISTA (CA-SPNM follows a similar pattern). While the data matrix $X$ is distributed among processors in line 3, randomize sampling is used in line 6 to reduce the cost of the matrix-matrix and matrix-vector operations in line 7, which allows us to control on-node computation and communication costs. We then stack these results in the Gram matrices in line 8 and do an all-reduce operation every \textit{k} iterations at line 9. Blocks of data from the Gram matrices are selected in line 11 based on the data dimensions and recurrence updates happen at line 12.

% \todo[inline,color=red]{SS: you may want to change or remove the next paragraph.}
% Randomized sampling is also used to in line 9 of both communication-avoiding algorithms which allows us to control on-node computation and communication costs by randomly sampling a few columns of the Gram matrix on every processor. % based on the number of selected columns and the flops cost associated with that.
% Operating on a subset of columns at each iteration can affect converge rates and this the parameter \textit{b} and the sampling size should be tuned.
% %%%%% end

\captionsetup[table]{name=Algorithm}
\setcounter{table}{4}
\begin{table}
    \caption{Distributed Implementation of CA-SFISTA}
    \centering
	\label{tab:pseudo-fista}
	\begin{tabular}{|l|p{2cm}|}
% 		\hline
        
%         \multicolumn{2}{|c|}{Algorithm 1 FISTA Algorithm}\\
		\hline
		1. INPUT: k, T, b, $\lambda$, $t$ and training dataset X, y\\
		2. INITIALIZE: $w_0=0$\\
        3. Distribute X column-wise on all processors so each\\
        \quad  processor roughly has the same amount of non-zeros\\
        4. for i=0,...,$\frac{T}{k}$ do \\
        5. \quad for j=1,...,k do on each processor\\
        6.\quad \quad \quad  Randomly select b percent of columns of X and rows of y\\
        7.\quad \quad \quad compute $XX^T$ and $XY$\\
        8.\quad \quad \quad stack the results in Gram matrices\\
        9. \quad \textbf{All-reduce the Gram matrices}\\
        10.\quad for j=1,...,k do on each processor\\
        11.\quad \quad compute gradient using blocks of Gram matrices\\
        12.\quad \quad update $w$ , $w \leftarrow S_{\lambda t_j}(w-t_j \nabla f(w))$\\
        13. RETURN $w_T$\\
        \hline

	\end{tabular}
\end{table}
\captionsetup[table]{name=Table}
\setcounter{table}{0}

\subsection{Cost of Communication-Avoiding Algorithms}
\label{analysis:ca}
We discuss the computation, storage, and communication costs of CA-SFISTA and CA-SPNM in following theorems and show that these algorithms reduce latency costs while preserving both bandwidth and flops costs.\\
\textbf{Theorem 3.} $T$ iterations of CA-SFISTA on $P$ processors over the critical path has the following costs:
F = $O(\frac{Td^2bn}{P})$ flops, W = $O(Td^2log P)$  words moved, L = $O(\frac{T}{k} log P)$ messages and M = $O(\frac{dn}{P}+kd^2)$ words of memory.

\textit{Proof.} CA-SFISTA computes $G$  which requires $O(\frac{kd^2bn}{P})$ operations, communicates $O(kd^2log P)$  words,  and requires $O(log P)$ messages. Computing $R$ requires $O(\frac{kdbn}{p})$ operations and communicates $O(kdlog P)$ words with $O(log P)$ messages. Then the algorithm computes the gradient and solves the minimization problem redundantly on all processors using a soft thresholding operator which requires $O(kd^2)$ operations for computing the gradient and $O(kd)$ for the soft thresholding operator without any communication. The vector updates on $w_{ik+j}$ can be done without any communication. Since the critical path occurs every \textit{k} iteration then the algorithm costs $O(\frac{Td^2bn}{P})$ flops, $O(Td^2log P)$  words and $O(Tlog P)$ messages. Each processor needs to store $G, R, v_{ik+j}, w_{ik+j}$, and $\frac{1}{P}$-th of $X$. Therefore, it needs $kd^2+sd+d+d+\frac{dn}{P}=O(\frac{dn}{P}+kd^2)$ words of memory.

% \todo[inline, color=green]{Topic paragraph! } 
\textbf{Theorem 4.} $T$ iterations CA-SPNM on P processors over the critical path has following costs:
F = $O(\frac{Td^2bn}{P}+\frac{Td^2}{\epsilon})$ flops, W = $O(Td^2log P)$  words moved, L = $O(\frac{T}{k} log P)$ messages and M = $O(\frac{dn}{P}+kd^2)$ words. \\
\textit{Proof.} CA-SPNM computes $G$ and $R$ and sends them to all processors. Each processors solves the minimization problem redundantly for $k$ iterations. At each inner iteration, the ISTA algorithm solves the minimization problem in $Q$ iterations. In order to get an $\epsilon$-optimal solution, the algorithm needs to run for $O(1/\epsilon)$ iterations; CA-SPNM requires $O(\frac{kd^2}{\epsilon})$ operations for this. A analysis similar to the proof to theorem 1 proves this theorem.

 Table \ref{tab:cost-table} shows the summery of costs for the CA algorithms compared to the classical algorithms. % It reduces the latency cost without changing the bandwidth and flop cost. 

\begin{table*}
	\centering
	\caption{Latency, floating point operations, memory, and bandwidth costs for different algorithms; $d$, $n$, $P$, $k$, and $\epsilon$ represent number of rows, number of columns, number of processors, step parameter, and the accuracy of the first order inner solver.}
	\label{tab:cost-table}
	\begin{tabular}{|c|c|c|c|c|p{1cm}|p{2cm}|p{1cm}|p{2cm}||p{2cm}|}
    \hline
		Algorithm & Latency cost (L)&Ops cost (F)&Memory cost (M)&Bandwidth cost (W)\\
        \hline
        SFISTA & $O(T log P)$&$O(\frac{Td^2bn}{P})$&$O(\frac{dn}{P})$&$O(Td^2log P)$\\ 
        \hline
        CA-SFISTA & $O(\frac{T}{k} log P)$&$O(\frac{Td^2bn}{P})$&$O(\frac{dn}{P}+kd^2)$&$O(Td^2log P)$\\ 
		\hline
        SPNM& $O(T log P)$ & $O(\frac{Td^2bn}{P}+\frac{Td^2}{\epsilon})$&$O(\frac{dn}{P})$ & $O(Td^2log P)$\\ 
		\hline    
        CA-SPNM& $O(\frac{T}{k} log P)$ &$O(\frac{Td^2bn}{P}+\frac{Td^2}{\epsilon})$ &$O(\frac{dn}{P}+kd^2)$ &$O(Td^2log P)$\\ 
		\hline
		
	\end{tabular}
\end{table*}

% \subsection{Implementation}

% \todo[color=green, inline]{Summery of results seems redundant. Cant you move parts of it that talks about Table II to section C as its concluding paragraph and remove the summery of results subsection.}

% \subsection{Summary of Results}
% \todo[inline, color=green]{This is what I recommend: Move sections E  and F to the next section IV. Section D should go to either IV or V}
% we reduce the communication for solving a regularized least square problem by a factor of $O(k)$ without changing the overall bandwidth and flop cost for $T$ iteration. The communication avoiding algorithm communicates more words every $k$ iteration, but it preserve the total bandwidth cost while keeping the convergence behavior intact. We unroll the update recurrences for fast iterative thresholding and stochastic proximal Newton method by a factor of k, then compute the stochastic Hessian and residual for s iterations beforehand and send it to all processors. Every processor updates optimization variable using a soft thresholding operator.

% \todo[inline,color=orange]{ MM: k and s are misused in the algorithm and in different parts of the paper. The inner unrolled loops should all be k since we are referring to CA methods as k-step methods.}

\section{Experimental Results}

This section presents experimental setup and performance results. The system setup and datasets as well as algorithm parameter settings are discussed first. We then show the effect of parameters \textit{k} and \textit{b} on the algorithms' convergence rate. Speedup and strong scaling properties of the communication-avoiding algorithms are demonstrated in the last section and are compared to the classical algorithms. 

% Also, we show that the parameter \textit{b} can be used to tune change the amount of computation on each processor without changing the stability too much.
\subsection{Methodology}

%    \todo[inline,color=green]{Still need to add your platform info and that you are using MPI. If you are limited to 64 processors by the cluster say that here. }

Table \ref{tab:datasets} shows the datasets used for our experiments. The datasets are from dense and sparse machine learning applications from  \cite{chang2011libsvm} and vary in size and sparsity. 
%   \textit{Architecture:} All convergence analysis experiments were performed in MATLAB version R2016b on a 2.5 GHz Intel Core i7 machine with 16GB of RAM. 
The algorithms are implemented in C/C++ using Intel MKL for (sparse/dense) BLAS routines and MPI for parallel processing. While use a dense data format for our theoretical analysis, we use the compressed sparse row format (CSR) to store the data for sparse datasets. Our experiments are conducted on the XSEDE Comet systems \cite{towns2014xsede}. %Since We tried to have a flat architecture in order to be fair in comparing communication costs. Therefore, for nodes less than 64, we have one task per processor. Also, the experiments that were executed on more than 64 nodes have more task per processor. For example, in order to run the algorithms on 256 processors, we had to have 64 nodes and 4 tasks per processors. 

\textit{Selecting $\lambda$:} $\lambda$ should be chosen based on the prediction accuracy of the dataset and can affect convergence rates. We tune $\lambda$ for so that our experiments have reasonable running time.  The final tuned value for  $\lambda$, for all experiments, is 0.1 for \textit{abalone} and 0.01 for \textit{susy} and \textit{covtype}. 
 
   \textit{Stopping criteria:} The stopping condition in CA-SFISTA and CA-SPNM could be set based on two criteria: \textit{(i)} The algorithms can execute for a pre-specified number of iterations, shown with \textit{T} in Algorithms \ref{tab:ca-fista-alg} and \ref{tab:ca-spnm-alg}. We use this stopping criteria for our experiments on strong scaling (section \ref{sec:strongscaling}) since the number of operations should remain the same across all processors; \textit{(ii)} The second stopping criteria is based on the distance to the optimal solution. We define a normalized distance to the optimal solution, called \textit{relative solution error} obtained via $\frac{\| \hat{w}-w_{op} \|}{\|w_{op}\|}$,
   where $\|w_{op}\|$ is the optimal solution found using Templates for First-Order Conic Solvers (TFOCS) \cite{becker2011templates}. TFOCS is a first order method where the tolerance for its stopping criteria is $10^{-8}$. Putting a bound on the stopping criteria enforces the algorithms to run until a solution close enough to optimal is achieved. The CA-SFISTA and CA-SPNM algorithms have to be changed in line 3 to include a \textit{while-loop} with the condition to exit when the relative solution error becomes smaller than a tolerance \textit{tol}. The \textit{tol} parameter for each dataset is chosen to provide a reasonable execution time. This stopping condition is used for the speedup experiments in section \ref{sec:speedup}. 

\begin{table}[h]
  \centering
 \begin{tabular}{ |p{2cm}|p{1cm}|p{1cm}|p{1cm}|p{1cm}|p{2.5cm}|  }
 %\hline
% \multicolumn{4}{|c|}{Country List} \\
 \hline
 Dataset & Row numbers & Column numbers & Percentage of nnz & Size (nnz)
\\
 \hline
% 2D Markov model & 3X & 11 & 3.52&  1.02\\
abalone & 4177 & 8 & 100\% & 258.7KB\\
susy & 5M & 18 & 25.39\% & 2.47GB\\
covtype &  581,012 & 54 & 22.12\%& 71.2MB\\
 \hline
\end{tabular}
%  \vspace{-0.10in}
  \caption{ The datasets.}
%    \todo[inline, color=green]{MM: Is matrix size the number of nonzeros in bytes? Cite the matrix source}
  \vspace{-0.1in}
  \label{tab:datasets}
\end{table}

  % Table \ref{tab:parameters} shows the parameters used for each dataset. These parameters give us an reasonable execution time to compare our results between different methods.
    
% \todo[inline,color=red]{SS: I should put a Table here for parameters. Columns: dataset, lambda, b, tol}
%%%%%%%%%%%%%%
% \begin{table}
%   \centering
%  \begin{tabular}{ |p{2cm}|p{1cm}|p{1cm}|p{1cm}|p{1cm}|p{2.5cm}|  }
%  \hline
%  Dataset & $\lambda$ & tol & step size & \textit{b} \\
%  \hline
% abalone & 0.1 & 0.1 & 0.1 & 10\%\\
% susy & 0.01 & 0.1 & 0.1 & 1\%\\
% covtype &  0.01 & 0.1 & 0.1& 1\%\\
%  \hline
% \end{tabular}
%   \caption{ Parameters used in evaluating performance of algorithms}
%   \label{tab:parameters}
% \end{table}

%%%%%%%%%%%%%%

\subsection{Convergence Analysis}
This section shows the effect of the sampling rate \textit{b} on convergence. Computation costs of each processor can be significantly reduced with a smaller $b$, however, very small sample sizes can influence stability and convergence. We also demonstrate that changing $k$ in the \textit{k}-step formulations of SFISTA and SPNM does not affect their convergence and stability behavior compared to  the classical algorithms since the \textit{k}-step formulations are arithmetically the same as the original algorithms.  The \textit{susy} dataset also follows a similar analysis; we did not include its data due to space limitations. 

% Finally, we talk about speedup and scaling of CA-FISTA and CA-SPNM methods and show that their scalability and speedup has been improved on modern supercomputers.

\subsubsection{Effect of \textit{b} on Convergence}
\begin{figure*}[!htb]
  \centering
  \subcaptionbox{abalone}[.4\linewidth][c]{%
    \includegraphics[width=.34\linewidth]{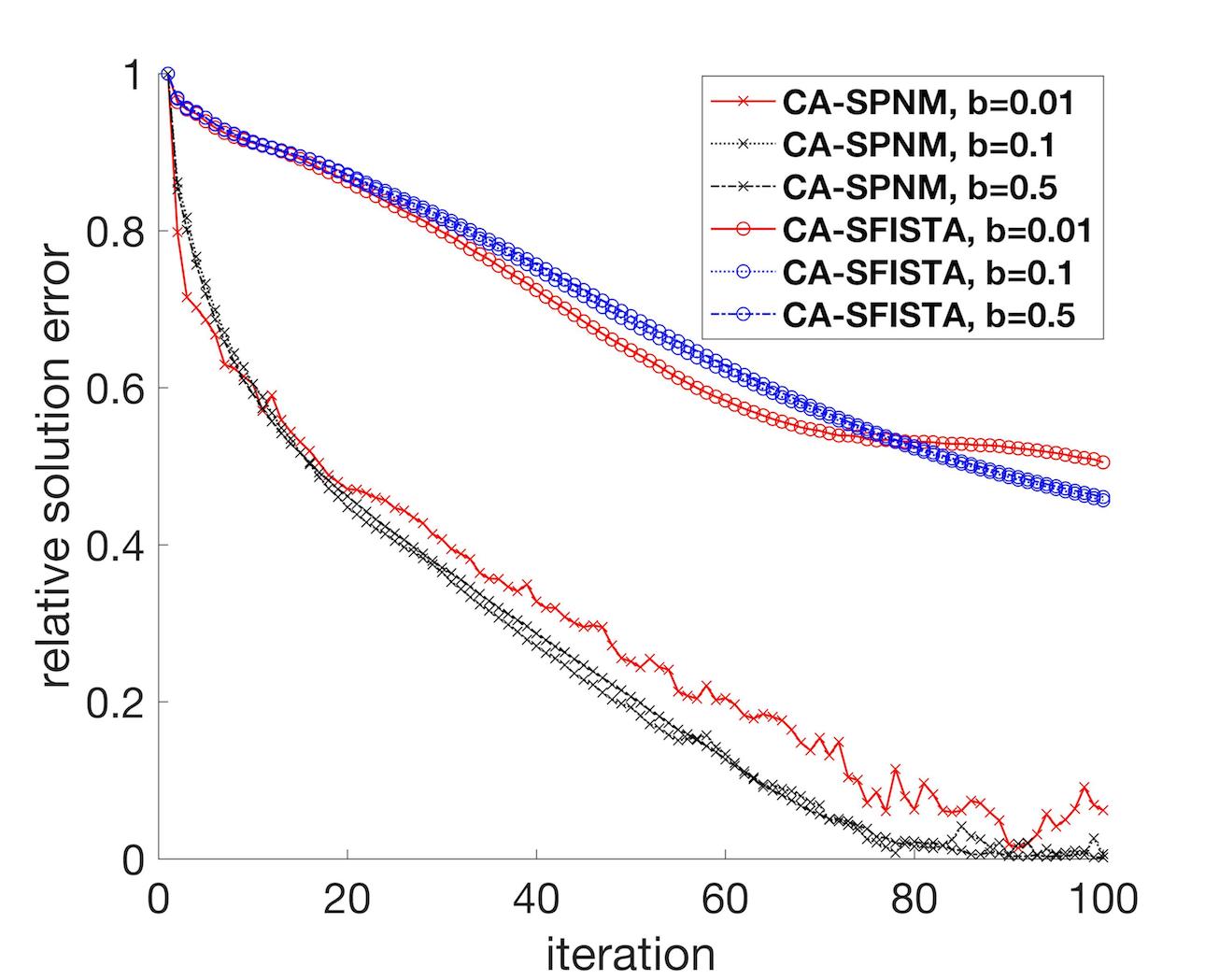}}\quad
  \subcaptionbox{covtype}[.4\linewidth][c]{%
    \includegraphics[width=.34\linewidth]{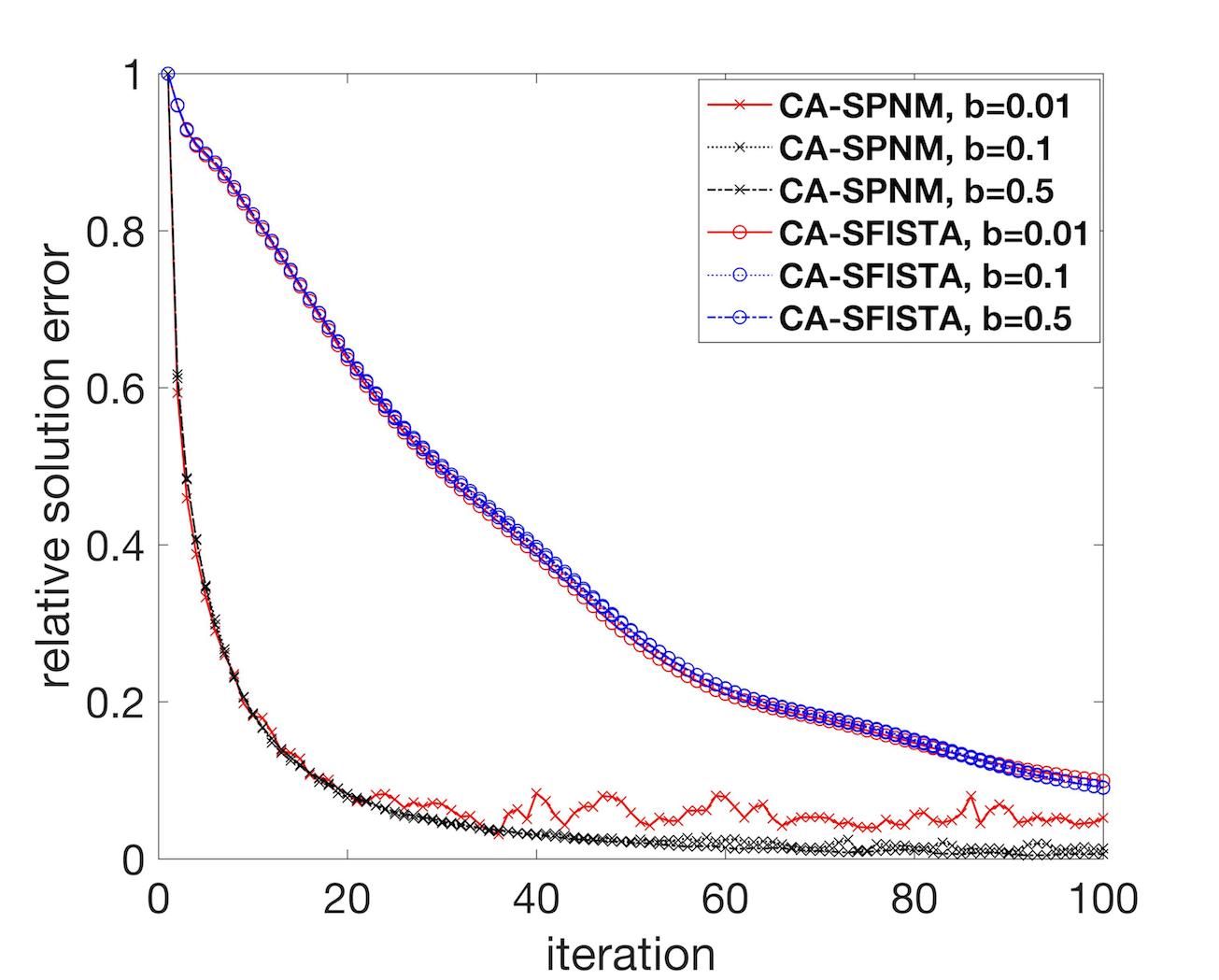}}\quad
      \caption{Effect of $b$ on convergence and stability for \textit{abalone} and \textit{covtype} datasets; $k$ is set to 32.}

  \label{fig:convergence_b}
\end{figure*}

Figure \ref{fig:convergence_b} shows the convergence behavior of CA-SFISTA and CA-SPNM for different sampling rates. The figure shows that for the same parameters, CA-SPNM converges faster than CA-SFISTA for all datasets. Smaller sampling rates reduce the floating point operation costs on each processor and as shown in Figure \ref{fig:convergence_b} the convergence rate does change for large values of \textit{b}. For example, we see that the residual solution error of CA-SFISTA follows the same pattern for both values of $b=0.01$ and $b=0.5$ for \textit{covtype}. However, if very few columns of the dataset is sampled then the gradient and Hessian may not represent a correct update direction. This gets worse as we get closer to the optimal solution, as shown in Figure \ref{fig:convergence_b} for $b=0.01$ on both datasets. %Using a One way to overcome this issue is using a diminishing step size which could be very effective. 
%Using a diminishing step size which could reduce this effect. 
%Also, we should mention that for all datasets, CA-SPNM method converge faster that CA-SFISTA with similar parameters.
%
%We showed that increasing the value of \textit{b} barely changes the convergence. Since we are more concerned about the communication bottleneck of the algorithm, we choose a small value for \textit{b} for remaining experiments, so the computation cost could be comparable and significant. 
For these experiments we set \textit{k} to 32; following shows the effect of $k$ on convergence.

\subsubsection{Effect of \textit{k} on Convergence}
\begin{figure*}[!htb]
  \centering
  \subcaptionbox{abalone}[.4\linewidth][c]{%
    \includegraphics[width=.34\linewidth]{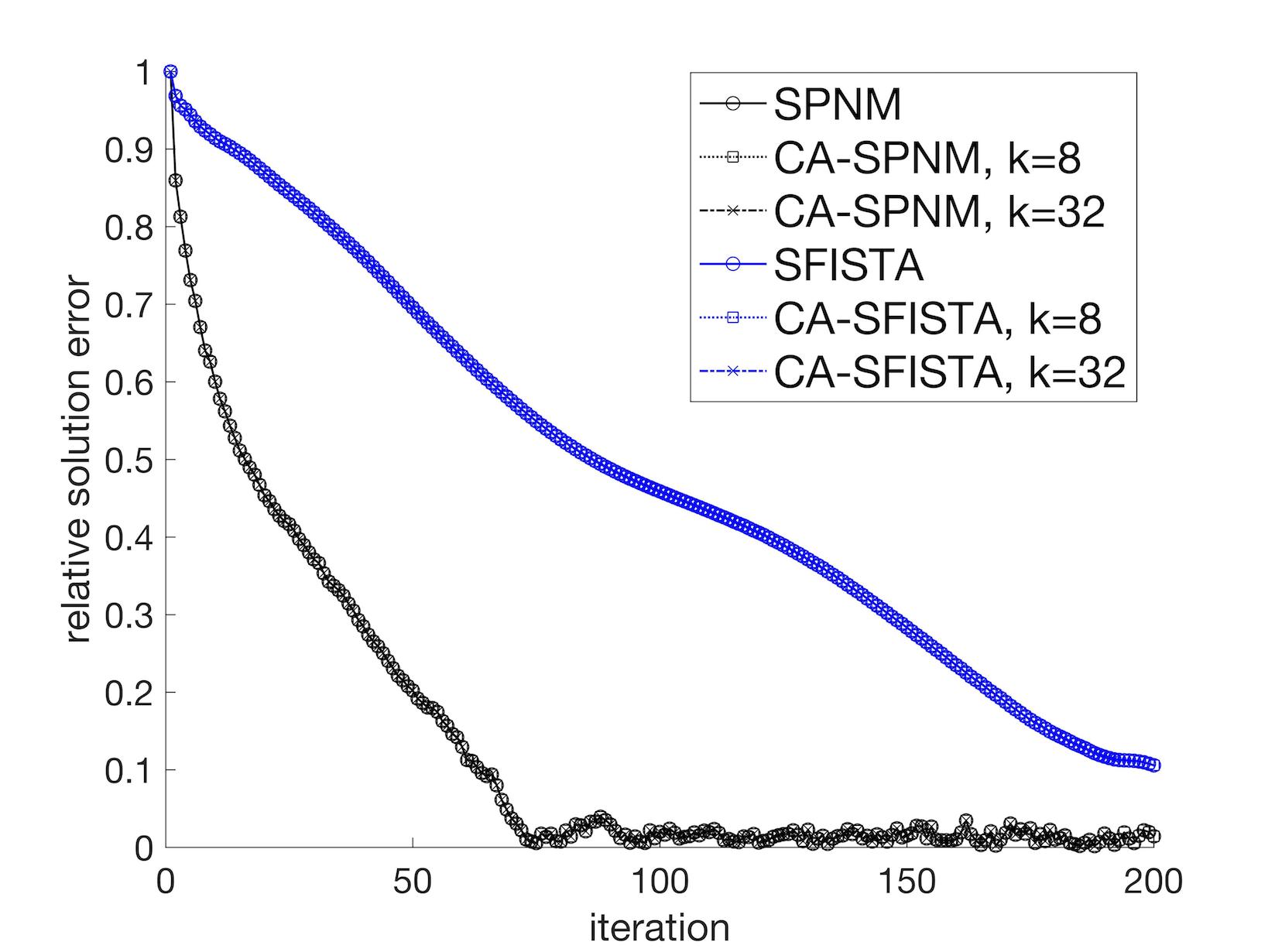}}\quad
  \subcaptionbox{covtype}[.4\linewidth][c]{%
    \includegraphics[width=.34\linewidth]{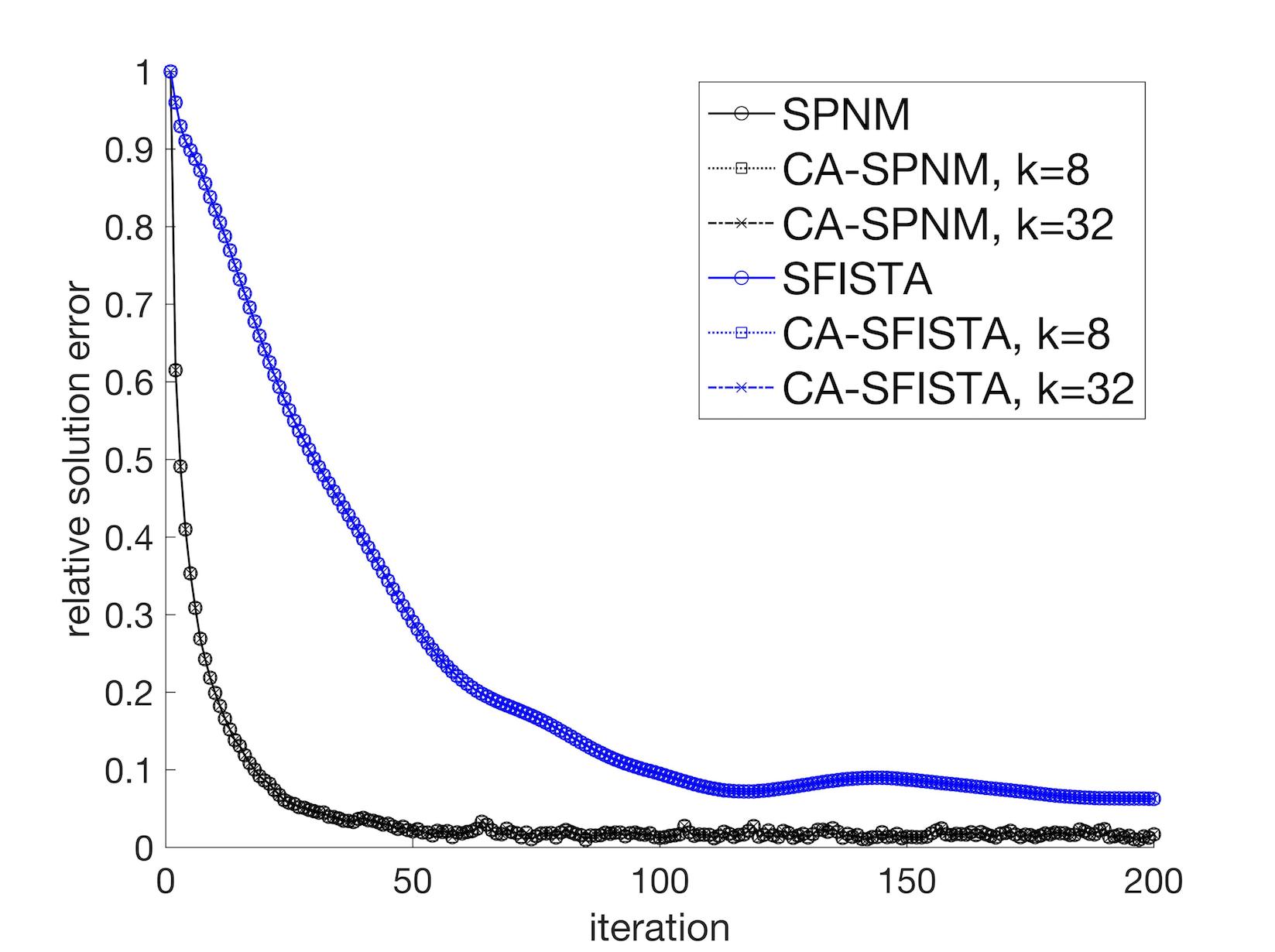}}\quad
      \caption{%Convergence behavior for two datasets. 
      Effect of $k$ on convergence and stability for \textit{abalone} and \textit{covtype}; $b=0.1$ for \textit{abalone} and $b=0.01$ for \textit{covtype}.}
      \vspace{-0.10in}
  \label{fig:convergence}
\end{figure*}

Figure \ref{fig:convergence} shows the convergence behavior of CA-SPNM and CA-SFISTA for two values of $k$ and shows the convergence rate of the classical SPNM and SFISTA algorithms. Since the CA derivations are arithmetically the same as the classical formulations, their convergence is also the same as SPNM and SFISTA. The experiments also demonstrate that changing $k$ does not affect the stability and residual solution error. We tested the convergence rate and stability behavior of the algorithms for up to $k=128$ and a similar trend was observed. For these experiments, $b$ is set to its best value.

% Also, one could increase the number of processors instead of choosing a small \textit{b} in order to bold the communication bottleneck of the algorithm.
 
\subsection{Performance Experiments}
Speedup and scaling results are presented in this section. Speedup results are obtained using the second stopping criteria and the tolerance parameter \textit{tol} is set to 0.1 for all the experiments. For the scaling experiments, we use the first stopping criteria with 100 iterations to execute the same number of operations across the experiments. The largest dataset \textit{susy} is executed on upto 1024 processors and the two smaller datasets \textit{abalone} and \textit{covtype} are executed on upto 64 and 512 processors respectively to report reasonable scalability based on size.

\subsubsection{Speedup}
\label{sec:speedup}
\begin{figure*}
  \centering
  \subcaptionbox{abalone}[.3\linewidth][c]{%
    \includegraphics[width=.28\linewidth]{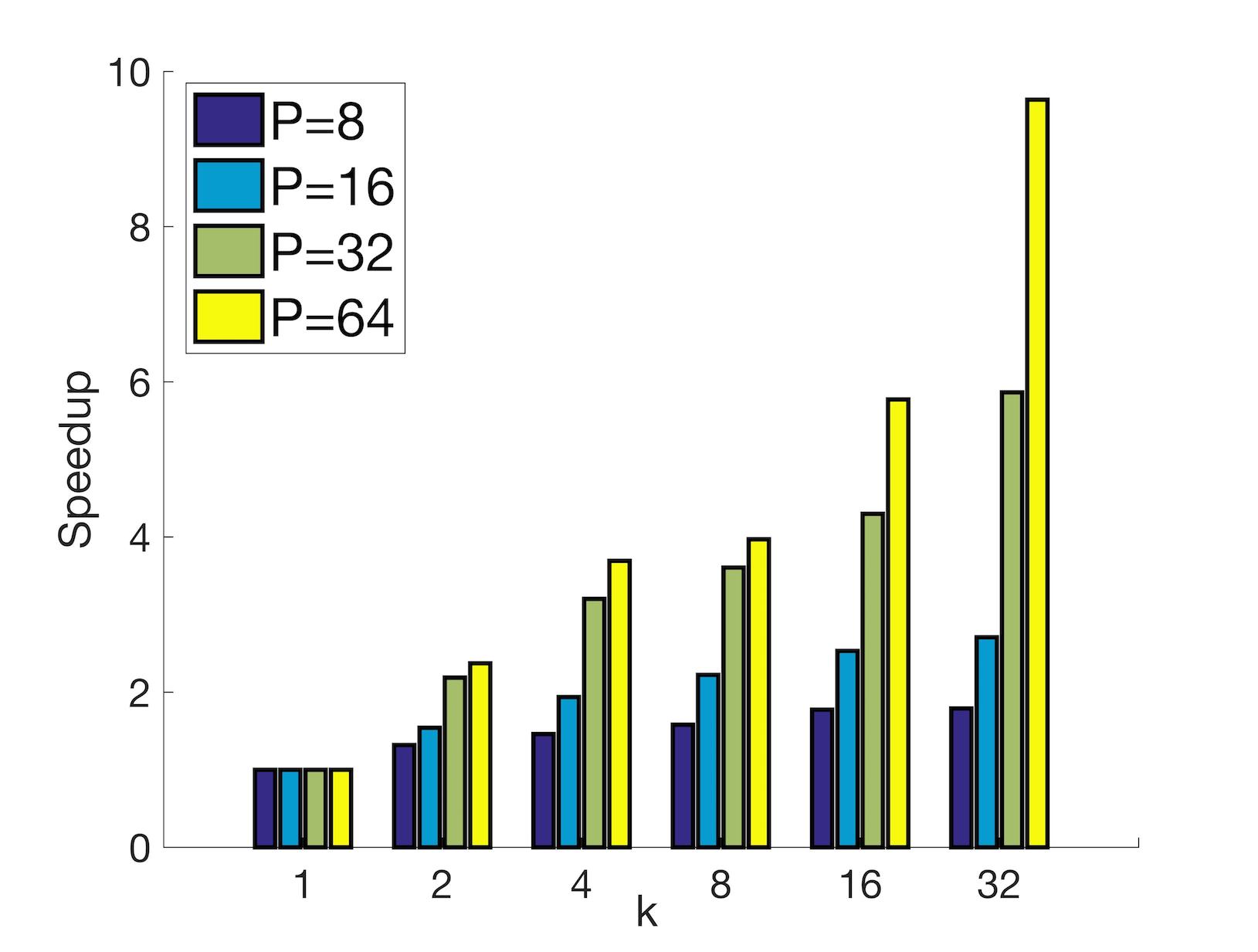}}\quad
  \subcaptionbox{covtype}[.3\linewidth][c]{%
    \includegraphics[width=.28\linewidth]{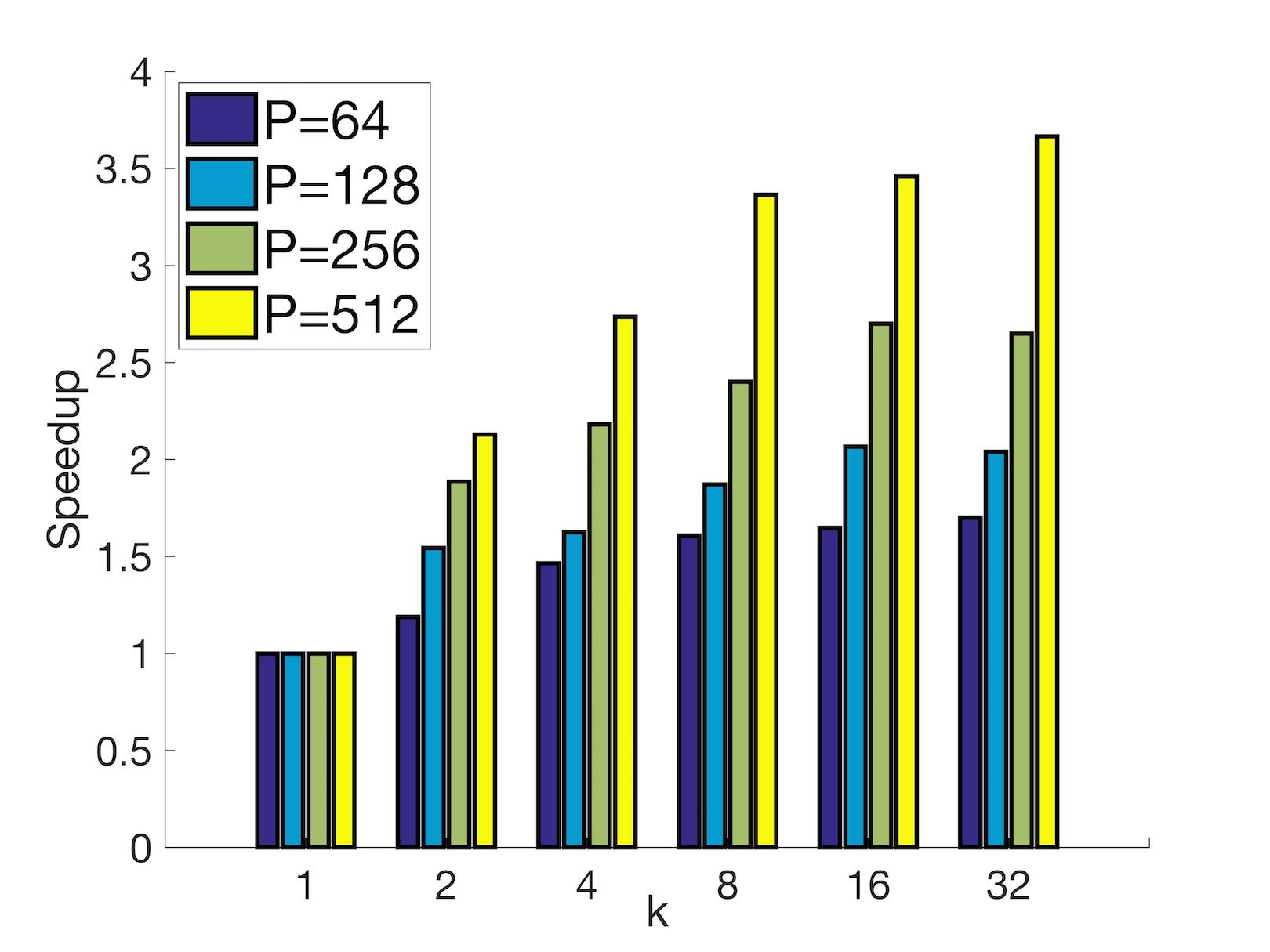}}\quad
  \subcaptionbox{susy}[.3\linewidth][c]{%
    \includegraphics[width=.28\linewidth]{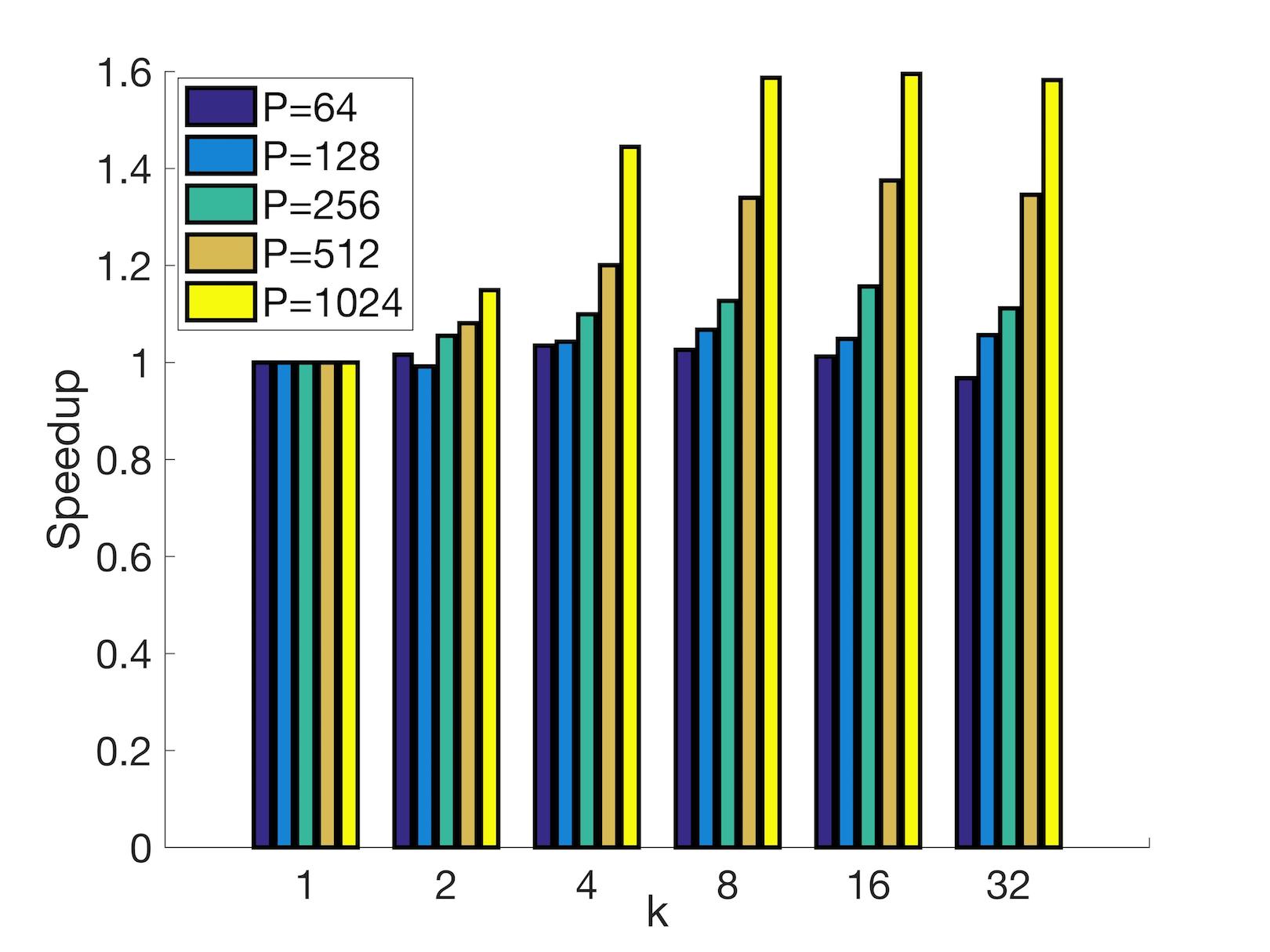}}

   \caption{Speedup for CA-SFISTA algorithm on three different datasets.}
   \label{fig:speedup_fista}
\end{figure*}

 Figure \ref{fig:speedup_fista} shows the speedup for all the datasets for different combinations of nodes (\textit{P}) and  \textit{k} for CA-SFISTA. All the speedups are normalized over SFISTA. At small scale, for example for $P=8$ and \textit{k}=32 for \textit{abalone} in Figure \ref{fig:speedup_fista}a a speedup of 1.79X is achieved, while for the same dataset we can get a speedup of 9.63X on 64 nodes. Increasing $k$ significantly improves the performance of \textit{abalone} since CA-SFISTA for a larger $k$ increases the ratio of floating point operations to communication for this relatively small data set on more processors. 

\begin{figure*}
  \centering
  \subcaptionbox{abalone}[.3\linewidth][c]{%
    \includegraphics[width=.28\linewidth]{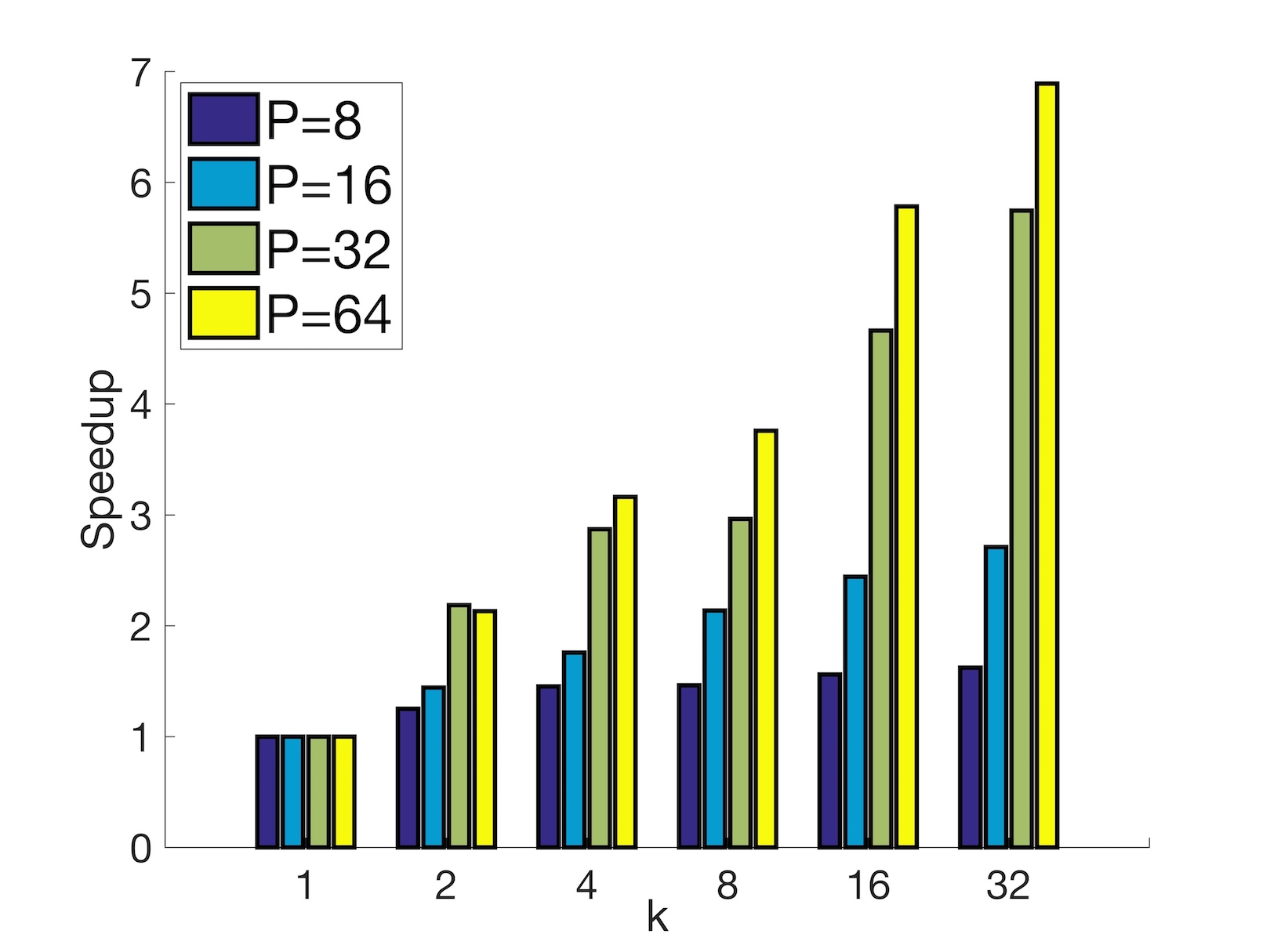}}\quad
  \subcaptionbox{covtype}[.3\linewidth][c]{%
    \includegraphics[width=.28\linewidth]{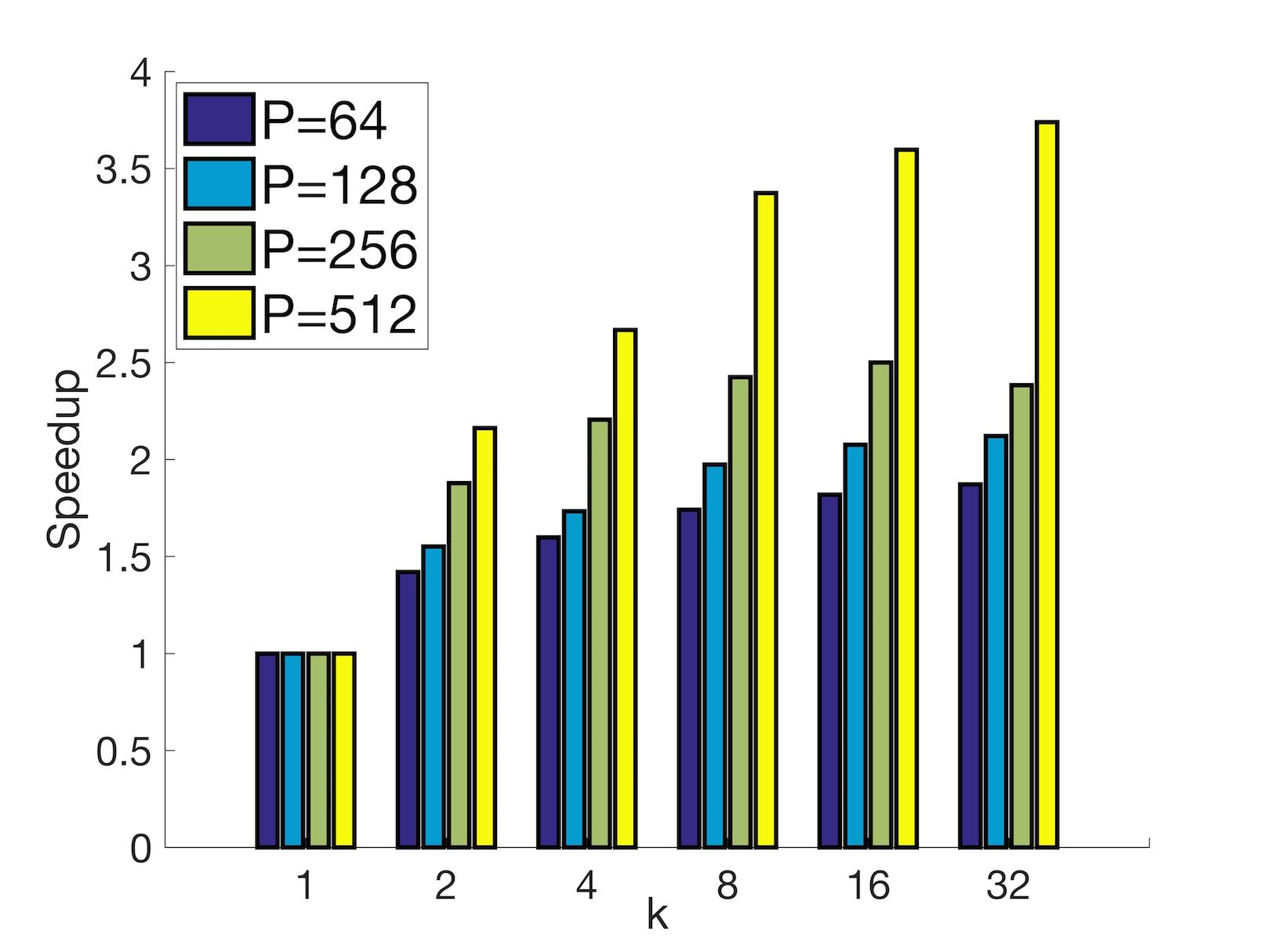}}\quad
  \subcaptionbox{susy}[.3\linewidth][c]{%
    \includegraphics[width=.28\linewidth]{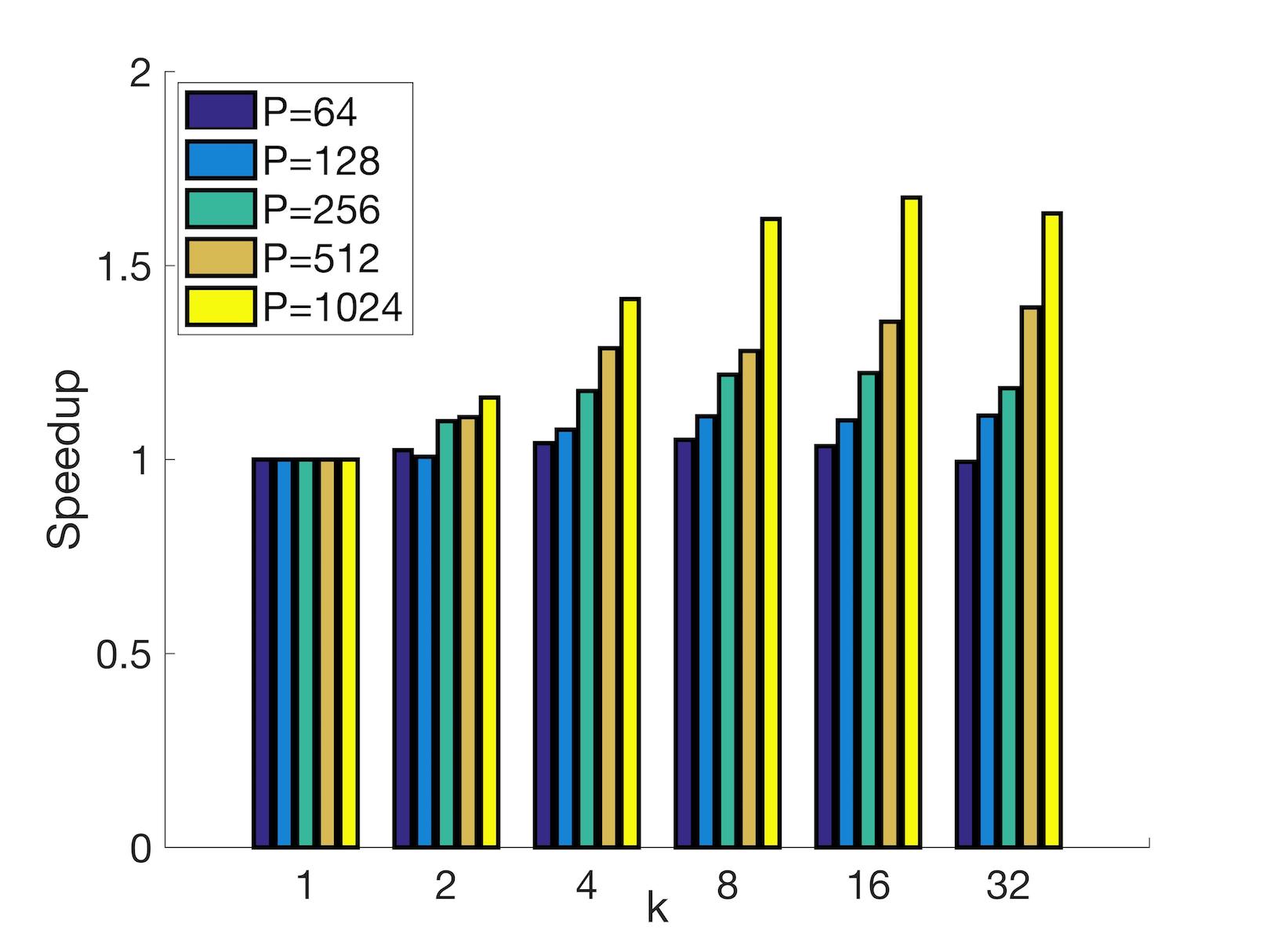}}

   \caption{Speedup for CA-SPNM algorithm on three different datasets.}
  \label{fig:speedup_spnm}
\end{figure*}

\begin{figure}
    \centering
    \includegraphics[width=1\linewidth]{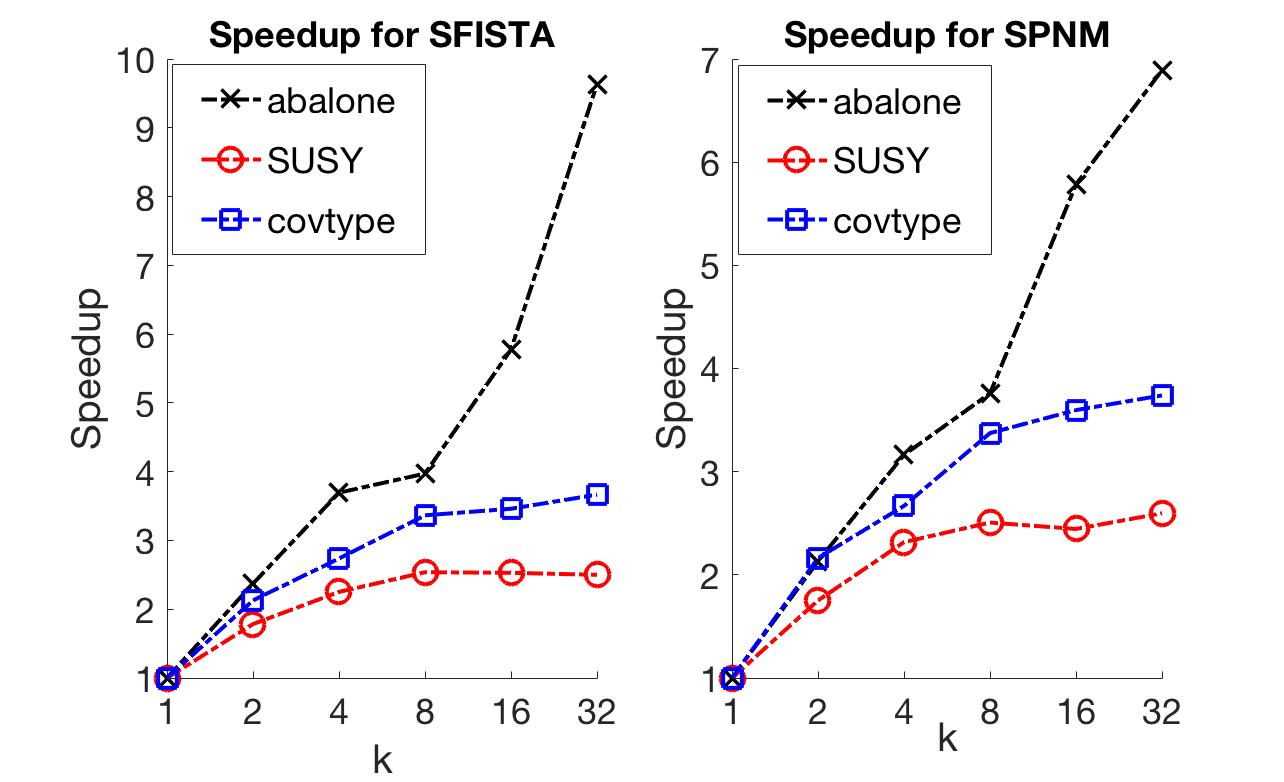}
    \caption{speedups on largest number of nodes.}
    \label{fig:speedups}
      \vspace{-0.15in}
\end{figure}

As shown in the figure, the performance of all datasets almost always improves when increasing the number of nodes or \textit{k}. For a fixed number of processors increasing $k$ reduces latency costs and communication  without changing bandwidth or arithmetic costs.  Figure \ref{fig:speedup_spnm} shows the speedup results for the CA-SPNM algorithm and follows a similar trend to CA-SFISTA. Figure \ref{fig:speedups} shows the speedup of CA-SFISTA and CA-SPNM algorithms on largest number of nodes for each dataset (64 for \textit{abalone}, 512 for \textit{covtype}, and 1024 for \textit{susy}). Increasing \textit{k} reduces the number of communicated  massages and latency costs. Therefore, the speedups improve as \textit{k} increases.
 
%Increasing the number of processors for the same $k$ shows that CA-SFITA scales well.%Therefore, we can expect greater speedups on more nodes, when latency becomes the dominant cost and we have a greater bandwidth.

%Figure \ref{fig:speedup_spnm} shows the speedup results for the CA-SPNM algorithm and follows a similar trend to CA-FISTA. The CA-SPNM dataset shows the same speedup for covtype and SUSY dataset but it has a smaller speedup compared to SFISTA. However, overall runtime for CA-SPNM outperform CA-SFISTA which is discussed in next section.
%Figure \ref{fig:speedups} shows the speedup of CA-SFISTA and CA-SPNM algorithms on largest number of nodes for each dataset (64 for \textit{abalone}, 512 for \textit{covtype}, and 1024 for \textit{susy}). Increasing \textit{k} reduce the number of communicated  massages and reduces latency costs. Therefore, the speedups improve as \textit{k} increases. %Based on the dataset dimension, \textit{k} and number of processors, we may experience a bandwidth limitation. For example, when CA-SFISTA is executed on covtype on 1024 nodes for \textit{k}=32, the number of words moved between processors at every \textit{k} iterations increases which makes the algorithm communicate more words. Notice that the overall bandwidth cost doesn't change. 
\begin{figure*}
  \centering
  \subcaptionbox{CA-SFISTA}[.4\linewidth][c]{%
    \includegraphics[width=.36\linewidth]{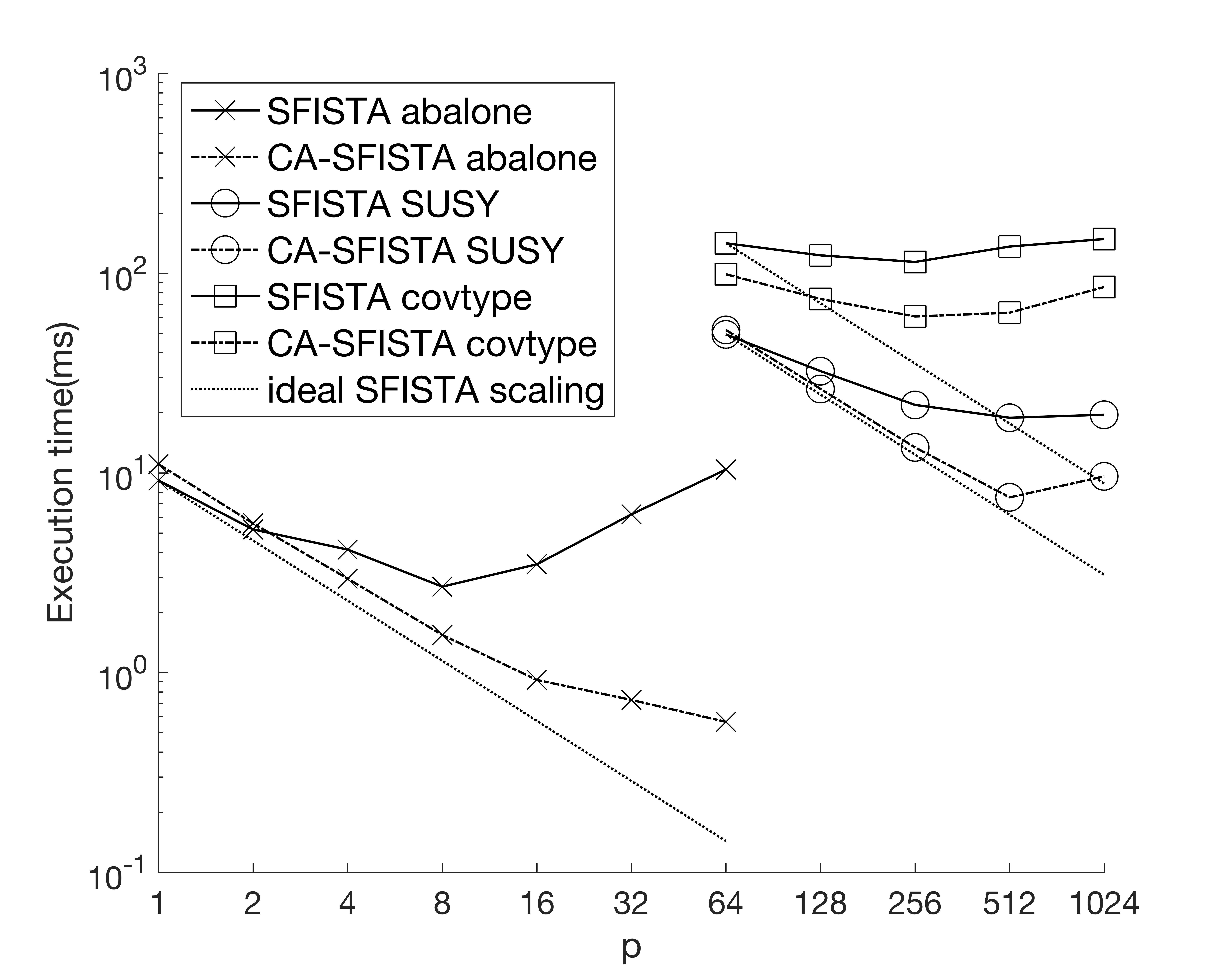}}\quad
  \subcaptionbox{CA-SPNM}[.4\linewidth][c]{%
    \includegraphics[width=.36\linewidth]{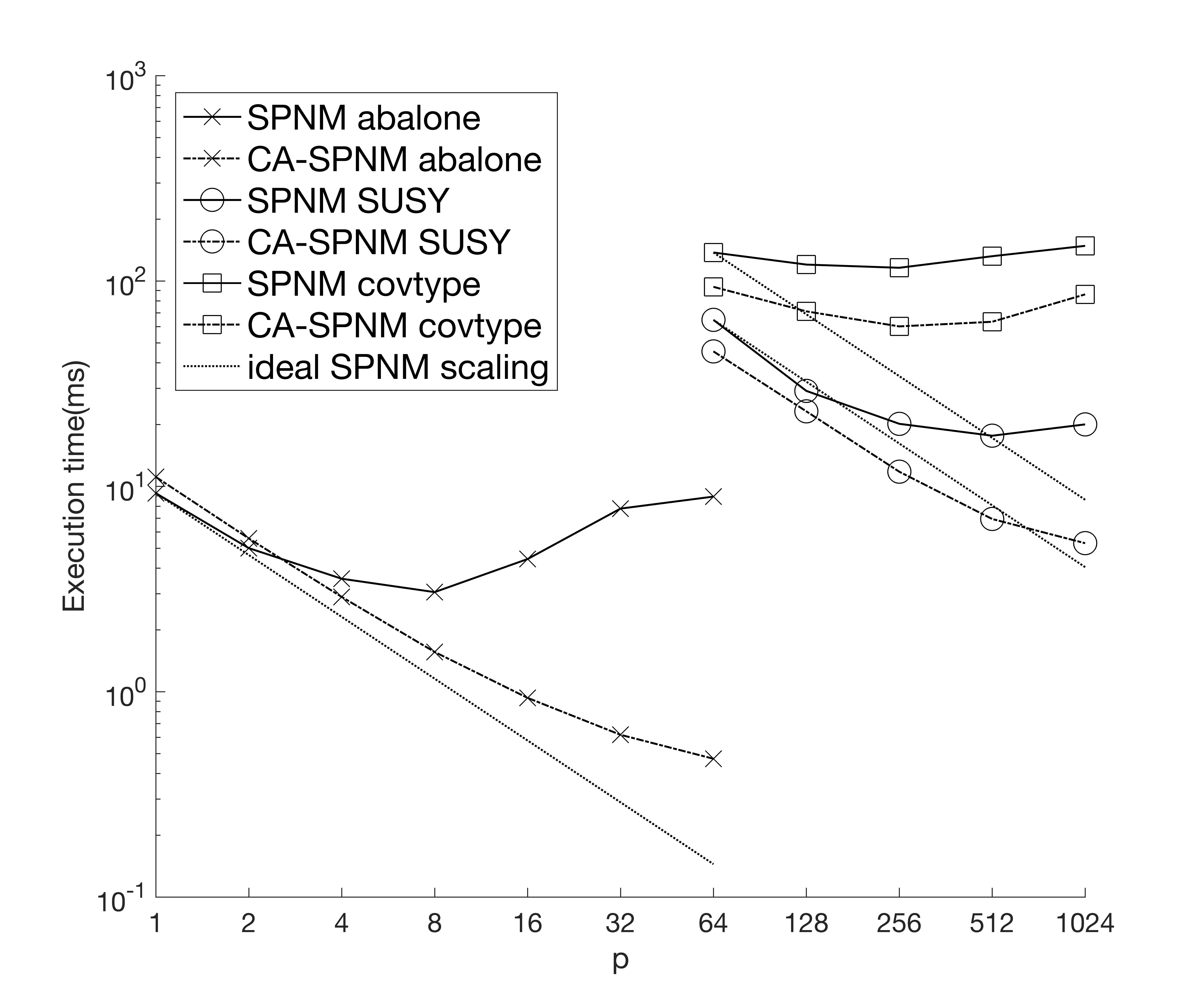}}\quad

   \caption{Comparing execution time of CA-SFISA and CA-SPNM with the classical algorithms.}
  \label{fig:strongscaling}
\end{figure*}

\subsubsection{Strong Scaling}
\label{sec:strongscaling}
Figure \ref{fig:strongscaling} (a) and Figure \ref{fig:strongscaling} (b) show strong scaling results for CA-SFISTA and CA-SPNM. The running times are the average over five runs for 100 iterations of the algorithm. The figures show the scaling behavior of the \textit{k}-step algorithms for \textit{k}=32 and compare it to SFISTA and SPNM. As shown in both figures, the classical algorithms do not scale well and their execution times increase when latency costs dominate. For example, for the \textit{abalone} dataset, SFISTA's running time increases on more than 8 processors, while CA-SFISTA continues to scale. The communication-avoiding methods are closer to ideal scaling. We intentionally added the $p=1024$ data point for the \textit{covtype} dataset to Figure \ref{fig:strongscaling} to show that the performance of the $k$-step algorithms is bounded by bandwidth. 
This is because increasing \textit{k} and the number of processors, increases the number of words to be moved in every \textit{k} iterations and performance is bounded by the bandwidth of the machine. If the processors had no bandwidth limitations then the algorithms would scale for all increasing \textit{k}.

\section{Conclusion}
Existing formulations of stochastic FISTA and proximal newton methods suffer from high communication costs and do not scale well on distributed platforms when operating on large datasets. We reformulate classical algorithms of SFISTA and SPNM to reduce inherent communication in their formulations. Our communication-avoiding algorithms leverage randomized sampling to overlap iterations and  reduce communication costs by a factor of \textit{k}. The proposed \textit{k}-step algorithms do not change the classical algorithms' convergence behavior and preserve the overall bandwidth and arithmetic costs. Our experiments show that CA-SFISTA and CA-SPNM provide  up to 10X speedup for the tested datasets on distributed platforms.
%\clearpage
\bibliographystyle{IEEEtran}
\bibliography{Sources/references.bib}

\begin{thebibliography}{10}
\providecommand{\url}[1]{#1}
\csname url@samestyle\endcsname
\providecommand{\newblock}{\relax}
\providecommand{\bibinfo}[2]{#2}
\providecommand{\BIBentrySTDinterwordspacing}{\spaceskip=0pt\relax}
\providecommand{\BIBentryALTinterwordstretchfactor}{4}
\providecommand{\BIBentryALTinterwordspacing}{\spaceskip=\fontdimen2\font plus
\BIBentryALTinterwordstretchfactor\fontdimen3\font minus
  \fontdimen4\font\relax}
\providecommand{\BIBforeignlanguage}[2]{{%
\expandafter\ifx\csname l@#1\endcsname\relax
\typeout{** WARNING: IEEEtran.bst: No hyphenation pattern has been}%
\typeout{** loaded for the language `#1'. Using the pattern for}%
\typeout{** the default language instead.}%
\else
\language=\csname l@#1\endcsname
\fi
#2}}
\providecommand{\BIBdecl}{\relax}
\BIBdecl

\bibitem{bottou2016optimization}
L.~Bottou, F.~E. Curtis, and J.~Nocedal, ``Optimization methods for large-scale
  machine learning,'' \emph{arXiv preprint arXiv:1606.04838}, 2016.

\bibitem{daubechies2004iterative}
I.~Daubechies, M.~Defrise, and C.~De~Mol, ``An iterative thresholding algorithm
  for linear inverse problems with a sparsity constraint,''
  \emph{Communications on pure and applied mathematics}, vol.~57, no.~11, pp.
  1413--1457, 2004.

\bibitem{beck2009fast}
A.~Beck and M.~Teboulle, ``A fast iterative shrinkage-thresholding algorithm
  for linear inverse problems,'' \emph{SIAM journal on imaging sciences},
  vol.~2, no.~1, pp. 183--202, 2009.

\bibitem{lee2012proximal}
J.~D. Lee, Y.~Sun, and M.~Saunders, ``Proximal newton-type methods for convex
  optimization,'' in \emph{Advances in Neural Information Processing Systems},
  2012, pp. 827--835.

\bibitem{dongarra2014applied}
J.~Dongarra, J.~Hittinger, J.~Bell, L.~Chacon, R.~Falgout, M.~Heroux,
  P.~Hovland, E.~Ng, C.~Webster, and S.~Wild, ``Applied mathematics research
  for exascale computing,'' Lawrence Livermore National Laboratory (LLNL),
  Livermore, CA, Tech. Rep., 2014.

\bibitem{ballard2014communication}
G.~Ballard, E.~Carson, J.~Demmel, M.~Hoemmen, N.~Knight, and O.~Schwartz,
  ``Communication lower bounds and optimal algorithms for numerical linear
  algebra,'' \emph{Acta Numerica}, vol.~23, pp. 1--155, 2014.

\bibitem{smith2016cocoa}
V.~Smith, S.~Forte, C.~Ma, M.~Takac, M.~I. Jordan, and M.~Jaggi, ``Cocoa: A
  general framework for communication-efficient distributed optimization,''
  \emph{arXiv preprint arXiv:1611.02189}, 2016.

\bibitem{recht2011hogwild}
B.~Recht, C.~Re, S.~Wright, and F.~Niu, ``Hogwild: A lock-free approach to
  parallelizing stochastic gradient descent,'' in \emph{Advances in neural
  information processing systems}, 2011, pp. 693--701.

\bibitem{zhou2017convergence}
F.~Zhou and G.~Cong, ``On the convergence properties of a $ k $-step averaging
  stochastic gradient descent algorithm for nonconvex optimization,''
  \emph{arXiv preprint arXiv:1708.01012}, 2017.

\bibitem{you2015svm}
Y.~You, J.~Demmel, K.~Czechowski, L.~Song, and R.~Vuduc, ``Ca-svm:
  Communication-avoiding support vector machines on distributed systems,'' in
  \emph{IEEE International Parallel and Distributed Processing Symposium},
  2015, pp. 847--859.

\bibitem{ppacksvm}
Z.~A. Zhu, W.~Chen, G.~Wang, C.~Zhu, and Z.~Chen, ``{P-packSVM}: Parallel
  primal gradient descent kernel svm,'' in \emph{Proceedings of the 2009 Ninth
  IEEE International Conference on Data Mining}, ser. ICDM '09.\hskip 1em plus
  0.5em minus 0.4em\relax Washington, DC, USA: IEEE Computer Society, 2009, pp.
  677--686.

\bibitem{chronopoulos1989efficient}
A.~T. Chronopoulos and C.~W. Gear, ``On the efficient implementation of
  preconditioned s-step conjugate gradient methods on multiprocessors with
  memory hierarchy,'' \emph{Parallel computing}, vol.~11, no.~1, pp. 37--53,
  1989.

\bibitem{CHRONOPOULOS1996623}
A.~Chronopoulos and C.~Swanson, ``Parallel iterative s-step methods for
  unsymmetric linear systems,'' \emph{Parallel Computing}, vol.~22, no.~5, pp.
  623 -- 641, 1996.

\bibitem{carson2015communication}
E.~C. Carson, \emph{Communication-avoiding Krylov subspace methods in theory
  and practice}.\hskip 1em plus 0.5em minus 0.4em\relax University of
  California, Berkeley, 2015.

\bibitem{carson2013avoiding}
E.~Carson, N.~Knight, and J.~Demmel, ``Avoiding communication in nonsymmetric
  lanczos-based krylov subspace methods,'' \emph{SIAM Journal on Scientific
  Computing}, vol.~35, no.~5, pp. S42--S61, 2013.

\bibitem{devarakonda2016avoiding}
A.~Devarakonda, K.~Fountoulakis, J.~Demmel, and M.~W. Mahoney, ``Avoiding
  communication in primal and dual block coordinate descent methods,''
  \emph{arXiv preprint arXiv:1612.04003}, 2016.

\bibitem{tibshirani1996regression}
R.~Tibshirani, ``Regression shrinkage and selection via the lasso,''
  \emph{Journal of the Royal Statistical Society. Series B (Methodological)},
  pp. 267--288, 1996.

\bibitem{schmidt2007learning}
M.~Schmidt, A.~Niculescu-Mizil, K.~Murphy \emph{et~al.}, ``Learning graphical
  model structure using l1-regularization paths,'' in \emph{AAAI}, vol.~7,
  2007, pp. 1278--1283.

\bibitem{mairal2009online}
J.~Mairal, F.~Bach, J.~Ponce, and G.~Sapiro, ``Online dictionary learning for
  sparse coding,'' in \emph{Proceedings of the 26th annual international
  conference on machine learning}.\hskip 1em plus 0.5em minus 0.4em\relax ACM,
  2009, pp. 689--696.

\bibitem{smith2015l1}
V.~Smith, S.~Forte, M.~I. Jordan, and M.~Jaggi, ``L1-regularized distributed
  optimization: A communication-efficient primal-dual framework,'' \emph{arXiv
  preprint arXiv:1512.04011}, 2015.

\bibitem{rendle2016robust}
S.~Rendle, D.~Fetterly, E.~J. Shekita, and B.-y. Su, ``Robust large-scale
  machine learning in the cloud,'' in \emph{Proceedings of the 22nd ACM SIGKDD
  International Conference on Knowledge Discovery and Data Mining}.\hskip 1em
  plus 0.5em minus 0.4em\relax ACM, 2016, pp. 1125--1134.

\bibitem{koh2007interior}
K.~Koh, S.-J. Kim, and S.~Boyd, ``An interior-point method for large-scale
  l1-regularized logistic regression,'' \emph{Journal of Machine learning
  research}, vol.~8, no. Jul, pp. 1519--1555, 2007.

\bibitem{fuller2011computing}
S.~H. Fuller and L.~I. Millett, ``Computing performance: Game over or next
  level?'' \emph{Computer}, vol.~44, no.~1, pp. 31--38, 2011.

\bibitem{culler1993logp}
D.~Culler, R.~Karp, D.~Patterson, A.~Sahay, K.~E. Schauser, E.~Santos,
  R.~Subramonian, and T.~Von~Eicken, ``Logp: Towards a realistic model of
  parallel computation,'' in \emph{ACM Sigplan Notices}, vol.~28, no.~7.\hskip
  1em plus 0.5em minus 0.4em\relax ACM, 1993, pp. 1--12.

\bibitem{alexandrov1995loggp}
A.~Alexandrov, M.~F. Ionescu, K.~E. Schauser, and C.~Scheiman, ``Loggp:
  incorporating long messages into the logp model—one step closer towards a
  realistic model for parallel computation,'' in \emph{Proceedings of the
  seventh annual ACM symposium on Parallel algorithms and architectures}.\hskip
  1em plus 0.5em minus 0.4em\relax ACM, 1995, pp. 95--105.

\bibitem{xiao2014proximal}
L.~Xiao and T.~Zhang, ``A proximal stochastic gradient method with progressive
  variance reduction,'' \emph{SIAM Journal on Optimization}, vol.~24, no.~4,
  pp. 2057--2075, 2014.

\bibitem{nitanda2014stochastic}
A.~Nitanda, ``Stochastic proximal gradient descent with acceleration
  techniques,'' in \emph{Advances in Neural Information Processing Systems},
  2014, pp. 1574--1582.

\bibitem{chang2011libsvm}
C.-C. Chang and C.-J. Lin, ``Libsvm: a library for support vector machines,''
  \emph{ACM transactions on intelligent systems and technology (TIST)}, vol.~2,
  no.~3, p.~27, 2011.

\bibitem{towns2014xsede}
J.~Towns, T.~Cockerill, M.~Dahan, I.~Foster, K.~Gaither, A.~Grimshaw,
  V.~Hazlewood, S.~Lathrop, D.~Lifka, G.~D. Peterson \emph{et~al.}, ``Xsede:
  accelerating scientific discovery,'' \emph{Computing in Science \&
  Engineering}, vol.~16, no.~5, pp. 62--74, 2014.

\bibitem{becker2011templates}
S.~R. Becker, E.~J. Cand{\`e}s, and M.~C. Grant, ``Templates for convex cone
  problems with applications to sparse signal recovery,'' \emph{Mathematical
  programming computation}, vol.~3, no.~3, pp. 165--218, 2011.

\end{thebibliography}
\end{document}